\documentclass[oneside,onecolumn]{article}

\usepackage{blindtext} 
\usepackage[sc]{mathpazo} 
\usepackage[T1]{fontenc} 
\usepackage{microtype} 

\usepackage[english]{babel} 

\usepackage[hmarginratio=1:1,top=32mm,columnsep=20pt]{geometry} 
\usepackage[hang, small,labelfont=bf,up,textfont=it,up]{caption} 
\usepackage{booktabs} 

\usepackage{lettrine} 

\usepackage{enumitem} 
\setlist[itemize]{noitemsep} 

\usepackage{abstract} 

\usepackage{titlesec} 
\renewcommand\thesection{\Roman{section}} 
\renewcommand\thesubsection{\roman{subsection}} 
\titleformat{\section}[block]{\large\scshape\centering}{\thesection.}{1em}{} 
\titleformat{\subsection}[block]{\large}{\thesubsection.}{1em}{} 

\usepackage{fancyhdr} 
\usepackage{titling} 

\usepackage{hyperref} 
\usepackage[utf8]{inputenc}
\usepackage[T1]{fontenc}
\usepackage{biblatex}
\usepackage{subcaption}

\usepackage{amsmath, amsthm, amsfonts, amssymb} 
\usepackage{mathrsfs}  
\usepackage{blkarray}
\usepackage{caption}
\usepackage{subcaption}
\usepackage{multirow}
\usepackage[switch]{lineno}
\usepackage{xcolor}
\usepackage{comment}
\usepackage{arydshln}

\usepackage{mathtools}
\DeclareUnicodeCharacter{2212}{-}

\title{\Large TEMPORAL CLUSTERING OF DISORDER EVENTS \\ DURING THE COVID-19 PANDEMIC} 
\author{%
\textsc{Gian Maria Campedelli}\thanks{Corresponding author} \\[0.5ex] 
\normalsize Department of Sociology and Social Research - University of Trento, Italy \\ 
\normalsize \href{mailto:gianmaria.campedelli@unitn.it}{gianmaria.campedelli@unitn.it} 
\and 
\textsc{Maria Rita D'Orsogna}\\[0.5ex] 
\normalsize Department of Computational Medicine - University of California, Los Angeles, CA 90095, USA \\ 
\normalsize Deparment of Mathematics - California State University at Northridge, CA 91330, USA
}
\date{April 22, 2021\\
\vspace{0.1cm}
\small This article has been published in PLOS One. Please cite it from here: \color{blue} \href{https://journals.plos.org/plosone/article?id=10.1371/journal.pone.0250433}{https://journals.plos.org/plosone/article?id=10.1371/journal.pone.0250433}} 
\renewcommand{\maketitlehookd}{%
\begin{abstract}
The COVID-19 pandemic has unleashed multiple public health, 
socio-economic, and institutional crises.  Measures taken to slow the spread of the virus
have fostered significant strain between authorities and citizens, leading to waves of 
social unrest and anti-government demonstrations. 
We study the temporal nature of pandemic-related disorder events as tallied by the ``COVID-19 Disorder Tracker" initiative 
by focusing on the three countries with the largest number of incidents, 
India, Israel, and Mexico. 
By fitting Poisson and Hawkes processes to the stream of data, 
we find that disorder events are inter-dependent and self-excite in all three countries. Geographic clustering confirms
these features at the subnational level, indicating that 
nationwide disorders emerge as the convergence of meso-scale patterns of self-excitation. 
Considerable diversity is observed among countries when computing correlations of events between subnational clusters; 
these are discussed in the context of specific political, societal and geographic characteristics.
Israel, the most territorially compact and where large scale protests were coordinated in response to government
lockdowns, displays the largest reactivity and the shortest period of influence
following an event, as well as the strongest nationwide synchrony. 
In Mexico, where complete lockdown orders were never mandated, 
reactivity and nationwide synchrony are lowest. Our work highlights
the need for authorities to promote local information campaigns 
to ensure that livelihoods and virus containment policies
are not perceived as mutually exclusive.\\
\vspace{0.5cm}

\textbf{Keywords:} Hawkes Process; Self-Excitability; India; Israel; Mexico; Disorder Events; Protests; Riots; Coronavirus
\end{abstract}
}

\begin{document}

\maketitle

\section{Introduction}

On March 11, 2020, the World Health Organization declared the coronavirus outbreak a global pandemic.  
No country has been spared from the far-reaching impacts of the ensuing viral disease, termed COVID-19, 
which touched and transformed most, if not all, aspects of societal living. In the absence of a widely available vaccine, and in light of the many unknowns posed by the pandemic, 
governments responded with a wide array of measures aimed at avoiding large scale public health crises with varying degrees of
resources, capacity, and resolve.  These included restrictions on travel and mobility, social distancing and quarantine 
requirements, imposing school and business closures,  implementing large scale testing, 
prohibiting large crowd gatherings. Periods of more or less stringent orders
alternated depending on the local ebbs and flows of the pandemic.
In parallel, governments also increased spending to reduce the risk of permanent damage to their economies
and to alleviate unemployment and poverty.
 Policymakers have been confronted with philosophical and moral questions centered on how to balance individual freedoms and preserving the economy
while protecting the lives of vulnerable populations, and on whether utilitarian-outcome or
kantian-deontological approaches should be taken \cite{HilsenrathEthicsEconomicsCOVID192020}. 
 In addition, especially at the onset of the pandemic, different jurisdictions adopted different strategies due to the limited scientific
understanding of COVID-19, and without being able to fully anticipate the medium-term health, social and political implications of their decisions. 
Sweden for example chose not to enter a lockdown while the rest of Europe did;  in the US emergency orders were implemented after non-uniform
threshold levels were met across the country, resulting in significant variability from county to county \color{black} \cite{BradyFragmentedUnitedStates2020}. \color{black}
Similarly, the recent announcement of the development of effective vaccines is marked by questions not only on how to optimally
scale up production but also how to distribute vaccines in an ethical manner \color{black} \cite{MillschallengesdistributingCOVID192021, MurielVaccineDistributionEquity2021}. \color{black}

While not perfect and difficult to evaluate, the above interventions for the most part helped prevent more widespread transmission and buffer worse
economic fallouts \color{black} \cite{AlfanoEfficacyLockdownCOVID192020, DelenNoPlaceHome2020, WelleniusImpactsUSStateLevel2020}. \color{black}  By the same token, they have provoked concerns over becoming drivers of 
economic uncertainty and deprivation \cite{PatelPovertyinequalityCOVID192020},  reinforcing
economic inequality \cite{DornCOVID19exacerbatinginequalities2020, 
GaleaCOVID19PandemicUnemployment2020}, 
engendering political distrust \cite{SibleyEffectsCOVID19pandemic2020}, and leading to 
long-term mental health consequences as a result of prolonged physical isolation and reduced human interaction 
\cite{GaoMentalhealthproblems2020a}.
Ultimately COVID-19 has caused what are poised to be long-lasting economic, financial, and social crises worldwide.

Traditional and social media have played a significant role in disseminating COVID-19 related information \color{black} \cite{Al-DmourInfluenceSocialMedia2020, ChenUnpackingblackbox2020}. \color{black} They 
have helped turn best practices into habits, and offered personalized networking opportunities in times of solitude. However, politicization of the pandemic,  
the sharing of non-vetted yet sensational or controversial findings or hypothesis, and algorithmic manipulation by platform hosts, 
have allowed for conspiracy theories and misinformation to spread, causing anxiety and discontent \color{black} \cite{PennycookFightingCOVID19Misinformation2020, TasnimImpactRumorsMisinformation2020, KouzyCoronavirusGoesViral2020} \color{black}. 

All these ingredients -- the emergence of a pandemic most were not prepared for, the difficulty for policymakers in make clear-sighted decisions that could be well received by a diverse set of stakeholders, the online spread of distorted information, personal fears and the need to connect to others -- have arguably combined to 
spark pandemic-related social unrest throughout the globe as manifest by the numerous protests, riots and disorders recorded
in many countries. Although these events have attracted 
media attention and ignited public debate, their characteristics and timing are
poorly understood.
Protests ignited by past outbreaks offer little insight into current societal unrest due to their 
comparatively limited spatio-temporal scope, and the lack of accurate data collection at the time they occurred.

In this paper we investigate the temporal distribution of civil disorders directly attributable to the 2020 coronavirus pandemic and, specifically, the temporal clustering and self-excitability of events. These ``contagion'' trends are recurrent in human activity and have been observed in criminal behavior \cite{ShortMeasuringModelingRepeat2009, KingHighTimesHate2013}, terrorist attacks 
\cite{LewisSelfexcitingpointprocess2012e, ChuangLocalalliancesrivalries2019a}, 
political shocks \cite{SimmonsGlobalizationLiberalizationPolicy2004, HouleDiffusionConfusionClustered2016}, violent conflicts \cite{BuhaugContagionConfusionWhy2008a, SchutteDiffusionpatternsviolence2011} and collective protesting and rioting \cite{Sullivancriticalmasscrowd1977,MidlarskyAnalyzingDiffusionContagion1978a, MyersDiffusionCollectiveViolence2000a}. We hypothesize they also characterize the current COVID-19 protests worldwide.

Our analyses focus on the three countries that were most hit by protests, riots, and violent events directly related to the COVID-19 pandemic, namely India, Israel, and Mexico,
as per data made available through the  ``COVID-19 \color{black}Disorder \color{black} Tracker'' (CDT) initiative \cite{ACLEDCOVID19DisorderTracker2020} curated by the ``Armed Conflict Location \& Event Data Project'' (ACLED)  \cite{RaleighIntroducingACLEDArmed2010}. While we recognize that peaceful protests are distinct from riots or episodes of violence against civilians, and that they
carry different features and theoretical implications \cite{GeschwenderCivilRightsProtest1968}, we do not discriminate among them in our analyses
due to their shared, high-level etiological trigger: the COVID-19 pandemic. 
Note that concurrent with the unfolding of the COVID-19 crisis are other episodes of civil unrest whose primary source is not pandemic-related, most
notably the Black Lives Matter protests that originated in Minneapolis in late May 2020 in the wake of George Floyd's murder and that 
persist at this time of writing.  ACLED does not tally these events, and we do not include them in our
work as these demonstrations arise primarily in response to racial injustice, although 
racial disparities in COVID-19 cases may have contributed to the unrest.

For each of the three countries we focus on,  we study the temporal clustering and self-excitability of pandemic-related
demonstrations on the national and subnational scales. We assume all events recorded by ACLED
in a given country are part of the same underlying stream of events, whereas on the local level 
we construct mutually exclusive geographical clusters through $k$-means clustering.
Temporal self-excitation is studied by applying the Hawkes process \cite{HawkesSpectraselfexcitingmutually1971a}, a stochastic
point process initially used to understand aftershocks in the vicinity of an earthquake epicenter, and later
adapted to finance, cell signaling, \color{black} and disease-spread. In relation to social phenomena, Hawkes process have also been applied to describe and analyze the distribution of violent events, such as terrorist attacks \cite{TenchSpatiotemporalpatternsIED2016g}, gun violence \cite{LoefflerGunViolenceContagious2018},  and gang-related crimes \cite{BrantinghamGangViolentCrime2020}. \color{black}
Hawkes processes are non-Markovian extensions of Poisson processes. The latter are memory-less
with events following a random temporal distribution, whereas  the likelihood of an event in a Hawkes process depends
on past ones, leading to clustering, memory effects, and self-excitability. These features are well-suited to study 
human behaviors that appear to be patterned in time.  We fit both Poisson and Hawkes processes to data within
each cluster to verify if and how temporal clustering emerges at various subnational scales. 

Our analysis reveals that temporal clustering of pandemic-related demonstrations is a common  
feature in all three countries. Despite variations in the temporal distribution of events 
and in the magnitude of self-excitability and reactivity across the three different contexts, Hawkes processes 
 better capture the underlying dynamics present in the data compared to Poisson processes.
Furthermore, we find that self-excitability at the national level appears to emerge
as the convergence of subnational, cluster-based self-excitatory events, rather than as the result 
of a meso-level stream of events. 
Our results highlight the interplay between the national and subnational socio-political discourse in the emergence
of large-scale disorders that 
may erupt in response to centralized decisions, with information easily channeled through social media and other communication networks,
but may also manifest locally, with the global source of discontent becoming particularized to local grievances and sources of tension.

The remainder of this paper is organized as follows. In the Background section we briefly review the 
protesting and rioting literature in other contexts. We also provide a synthetic account of the evolution of the pandemic in each of the three countries under investigation.
In the Materials and Methods section will describe the ACLED data we utilize and outline our methodology.  The statistical outcomes of our inferential models, 
along with additional analyses on the temporal characteristics of events in each country and cluster, are presented in the Results section.
Finally, in the Discussion and Conclusion section we discuss the relevance of our results and their broader significance. 

\noindent

\section{Background}
\label{Back}

COVID-19-related social unrest has been observed in several countries in the form of protests, riots, and other demonstrations,
often in response to virus-containment decisions imposed by governments. These decisions may have seemed too onerous, unfair, or 
the root cause of economic uncertainty. It is important to understand the characteristics of this unrest for two reasons:
one is preventative, as demonstrations always carry the risk of widespread contagion, given their often chaotic and crowded nature, 
and of devolving into uncontrolled, violent clashes with authorities and/or among protesters and counter-protesters. 
The other is more introspective, as these protests may be manifestations of latent societal discontent, predating the pandemic,
invigorated by newly mandated restrictive policies. 

Though unique in its global reach, the current pandemic does not represent a \textit{unicum} in terms of the social unrest it has caused.
History has repeatedly been marked by uprising and protests in response to social, political, and economic crises. In recent decades, 
social movements and collective action have mobilized citizens in demanding changes to political regimes, economic policies, and 
more respect for human rights. A vast academic literature has emerged as a result, to study why, how, when large scale peaceful and violent
demonstrations occur and what their repercussions may be \cite{ClementPeopleHistoryRiots2016}.
Sociological and social movement studies 
\cite{DellaPortaSocialMovementsIntroduction2020} have been central in describing various 
theoretical frameworks underlying social unrest. Political science studies have helped dissect the political conditions in which riots and protests
flourish and anticipate possible micro-- and macro--consequences such as conflicts, regime changes, and revolutions \cite{HaleRegimeChangeCascades2013}. Criminological studies have concentrated on 
understanding how deviant behavior can emerge in non-peaceful assemblies and in outlining related policing strategies 
\cite{GreerWePredictRiot2010, 
BaudainsGeographicpatternsdiffusion2013}. Psychological studies have helped unveil the behavioral and cognitive mechanisms occurring 
in individuals when engaged in collective action \cite{ThomasWhenWillCollective2014, Drurysocialidentitymodel2020}. 

The last decades have also witnessed the increased use of quantitative methodologies to study disorder events of various nature that have
been used across disciplines and that have helped uncover consistent patterns. At the end of the 1970s, T.J. Sullivan \cite{Sullivancriticalmasscrowd1977}
and M.I. Midlarsky \cite{MidlarskyAnalyzingDiffusionContagion1978a},
while belonging to distinct academic fields (social psychology and political science, respectively),
highlighted ``universal'' features in the size distribution and dynamical evolution of crowds.
M.I. Midlarsky, in particular, emerged as a strong advocate for the incorporation of quantitative frameworks 
within the study of collective social phenomena. In one of his seminal works, he demonstrated the presence of two underlying processes 
driving the urban disorders of 1966-67 in several US cities, diffusion and contagion, at a time when neither had been proven.
The same empirical techniques were later applied to the study of transnational terrorism \cite{MidlarskyWhyViolenceSpreads1980c}. 

The greater availability of data in recent years has vastly improved our understanding of societal unrest and has led to 
refined statistical and computational approaches.
Sources include law enforcement and institutional organizations, 
social media platforms and news agencies \cite{PoellSocialmediatransformation2014, TrottierSocialMediaPolitics2014, JostHowSocialMedia2018}. 
Among the most investigated scenarios are the 2005 Paris riots \cite{Berestyckimodelriotsdynamics2015, Bonnasse-GahotEpidemiologicalmodelling20052018},  the 2011 London riots \cite{BaudainsTargetChoiceExtreme2013, DaviesmathematicalmodelLondon2013}, political protests in 
Latin America \cite{CadenaForecastingSocialUnrest2015}  and the Arab Spring \cite{HussainWhatBestExplains2013}. In all cases non-random behaviors were detected such as clustering processes, cascading dynamics, and self-excitability.  Complementary theoretical propositions helped diagnose the causal mechanisms of the 
inferred patterns and include hierarchical patterns \cite{MidlarskyMathematicalModelsInstability1970, LiCoupContagionHypothesis1975}, rational choice \cite{DaviesmathematicalmodelLondon2013}, identity theory \cite{McCallIdentitiesInteractionsExamination1978}.

In the spirit of expanding the extant literature on human dynamics during disorders, we focus on pandemic-related
demonstrations in the three nations that exhibit the greatest number of events, India, Israel and Mexico. Although these countries vary greatly in terms of geographical, socio-political, and economic fabric, our data-driven study will show that self-excitability and clustering emerge across all three as fundamental features of pandemic-related protests and riots.

\subsection{Overview of the pandemic in select countries}

\subsubsection{India}
Currently, India is the country with the second-highest recorded number of COVID-19 
cases worldwide, with more than 10 million infections, corresponding to roughly 
12\% of the global count \cite{Donginteractivewebbaseddashboard2020}. 
The first case to be confirmed was a medical student 
returning from Wuhan, China to the state of Kerala who tested positive 
on January 30$^{\rm th}$ 2020 \cite{AndrewsFirstconfirmedcase2020}. 
 India's first fatality was reported on March 12$^{\rm th}$, amidst growing numbers of infections. 
On March 22$^{\rm nd}$,  Prime Minister Narendra Modi 
called for a 14-hour voluntary national lockdown, which became 
mandatory  three days later, when only essential services were allowed to remain open.
On April 14$^{\rm th}$ the lockdown was extended to May 3$^{\rm rd}$ and on  
April 29$^{\rm th}$ India's death toll reached one thousand.
On May 1$^{\rm st}$, the lockdown was extended
for two more weeks \color{black} although some flexibility was allowed, 
depending on the local spread of the virus. \color{black} Other economic, financial and social 
measures were introduced as lockdown restrictions were extended through the end of May. 
Starting in June, containment policies were progressively 
eased. On August 30$^{\rm th}$, the country recorded the highest number of new cases worldwide, in excess of 78,000 \cite{BhardwajIndiasetsglobal2020}; 
this record was shattered on September 12$^{\rm th}$ when India tallied more than 95,000 infections.  
Schools were partially reopened at the end of summer and
on September 30$^{\rm th}$ states and Union Territories were allowed local autonomy on certain pandemic-related matters.

Protests against the government’s management of the COVID-19 crisis have been widespread and sustained.
Migrant workers unable to return home due to lockdown orders staged protests at railway stations, 
health workers went on strike over the lack of protective equipment and adequate pay, 
agricultural workers demanded payment of lockdown wages and protested the lack of aid despite government promises, 
students demonstrated against holding University admission exams and
requested they be deferred due to infection risks and the difficulty of reaching exam centers.
General strikes were organized to rally against poverty and unemployment triggered by COVID-19.  
On some occasions, episodes of state repression and violent enforcement of lockdown orders occurred
as an attempt to quell these disorders. 
Long standing conflicts tied to identity politics, immigration, secessionist movements and boundary disputes
have also been impacted by COVID-19. Massive protests had engulfed the country in 2019 and early 2020
in response to the Citizenship Amendment Act (CAA) which would grant citizenship rights 
on the basis of religion \color{black} \cite{PokharelProtestsIndiaNew2019} \color{black}. Political and separatist demonstrations, including those related to the CAA,
declined following lockdown orders but reignited in June as restrictions eased, adding to pandemic-related
disorders. 
In India 50$\%$ of the total population had internet access in 2020 \color{black} \cite{DatareportalDigital2020India2020}. \color{black}

\subsubsection{Israel}
Israel recorded its first confirmed COVID-19 case on February 21$^{\rm st}$ 2020, 
after the return of a female citizen quarantined on the Diamond Princess in Japan \cite{LastfirstwaveCOVID192020}. 
Authorities soon issued a 14-day isolation policy for citizens who had visited South Korea or Japan; 
social distancing and related measures were imposed on March 11$^{\rm th}$, a ban on public gatherings of more than 10 people 
followed on March 15$^{\rm th}$. Finally, on March 19$^{\rm th}$, a 
state of emergency was declared and a national lockdown imposed.
A contact-tracing program was approved on March 16$^{\rm th}$,  allowing the 
Israeli Security Agency to track the mobility of individuals through mobile phone data. 
The program quickly sparked controversy and nationwide protests over the invasion of privacy 
and the overreaching surveillance of citizens. It was halted at the end of April. 
Additional restrictions were imposed between the end of March,
in anticipation of the Passover Seder festivity on April 8$^{\rm th}$ 
including a travel ban and the creation of a restricted zone in Bnei Brak, an ultra-Orthodox town
east of Tel Aviv with one of the highest rates of coronavirus cases in the country. 
Restrictions were eased between the end of April and the beginning of May. Retail stores 
were allowed to open on April 24$^{\rm th}$; school reopenings took place between May 3$^{\rm rd}$ and May 19$^{\rm th}$. 
In July, in response to steadily increasing ``second wave''  
infections, the Knesset reauthorized the contact-tracing program, once more igniting civil liberties organizations; 
new social distancing rules were also imposed. A ``traffic light'' monitoring plan was announced on August 31$^{\rm st}$
whereby a color representative of risk levels would be assigned to all Israeli cities \color{black} \cite{Jaffe-HoffmanRedlightgreen2020} \color{black}. Each color would be associated to more or less
restrictive rules. On September 6$^{\rm th}$ schools were closed
and night-time curfews were imposed on forty high risk communities, including nine in Jerusalem, affecting approximately 1.3 million people. 
A new national lockdown was issued for September 18$^{\rm th}$,  concurrent with the Jewish High Holy Days, 
and further restrictions announced on September 23$^{\rm rd}$.
On October 18$^{\rm th}$ some 
restrictions were lifted, with further reopenings announced throughout November 2020, although some high impacted
communities remained under lockdown.  Finally, a ``third wave'' of infections
was accompanied by a third nationwide lockdown imposed on December 27$^{\rm th}$. 

Protests in Israel have been common throughout the pandemic, with demonstrations against 
the government's handling of the crisis, and the alleged corruption of Prime Minister Benjamin Netanyahu. 
After restrictions were placed to curb demonstrations,  
nationwide protests were held on October 3$^{\rm rd}$. Clashes with police, rock-throwing
and confrontations between protesters and anti-protesters were recorded\color{black} \cite{ShpigelClashesarrestshundreds2020}\color{black}.
In Israel 84$\%$ of the total population had internet access in 2020 \color{black} \cite{DatareportalDigital2020Israel2020}. \color{black}

\subsubsection{Mexico}
On February 28$^{\rm th}$ 2020, Mexico confirmed its first COVID-19 cases as three men who had traveled to Bergamo, Italy; 
the first COVID-19 related death was recorded on March 18$^{\rm th}$. 
On March 23$^{\rm rd}$ President Andr\`es L\'opez Obrador unveiled a national campaign to promote social distancing, self-isolation,
and other measures to contain the spread of the virus. 
On March 30$^{\rm th}$, Mexico declared a state of health emergency, suspending all nonessential activities
for a month. Plans for gradual reopening were announced on May 18$^{\rm th}$, with the mayor of Mexico City presenting 
a blueprint for a ``new normality'' in the capital on  May 20$^{\rm th}$. On June 1$^{\rm st}$ the government imposed a traffic light
monitoring plan at the state level, similar to the one described for Israel.  Cases steadily increased starting from the
end of May leading to overcrowded hospitals. 
On July 30$^{\rm th}$ it was announced that governors who altered the traffic light status of their states would face criminal sanctions.
Record new cases of more than 10,000 new infections 
were recorded on November 25$^{\rm th}$. 

A few weeks into the pandemic, several media outlets began \color{black}reporting \color{black} that drug cartels were distributing food and medical supplies to citizens;
over time criticisms mounted on government's handling of the pandemic, the too modest spending on public heath and measures to help the economy, 
the many deaths in Mexico City, inadequate testing, medical personnel, waste disposal, and possible concealment of the real number of COVID-19 cases and fatalities. 
The president has been accused of having resisted full lockdown measures despite rising cases, to keep the economy open, 
drawing criticism from the WHO and the IMF. At this time of writing, Mexico is still among the most affected countries, 
with more than one fifth concentrated in its capital.

Labor groups have protested the government’s inadequate support of the population during the pandemic, 
including those who work in the informal economy; citizen groups demanded the resignation of President L\'opez Obrador
for his handling of the pandemic.  Other forms of pandemic-related disorders include
street vendors protesting closures of their businesses, residents protesting over unsubstantiated rumors of COVID-19 inducing
substances being spread by drones, against checkpoints that limited access to certain cities, and over mask-wearing.

\begin{figure}[!hbt]
    \centering
    \includegraphics[scale=0.6]{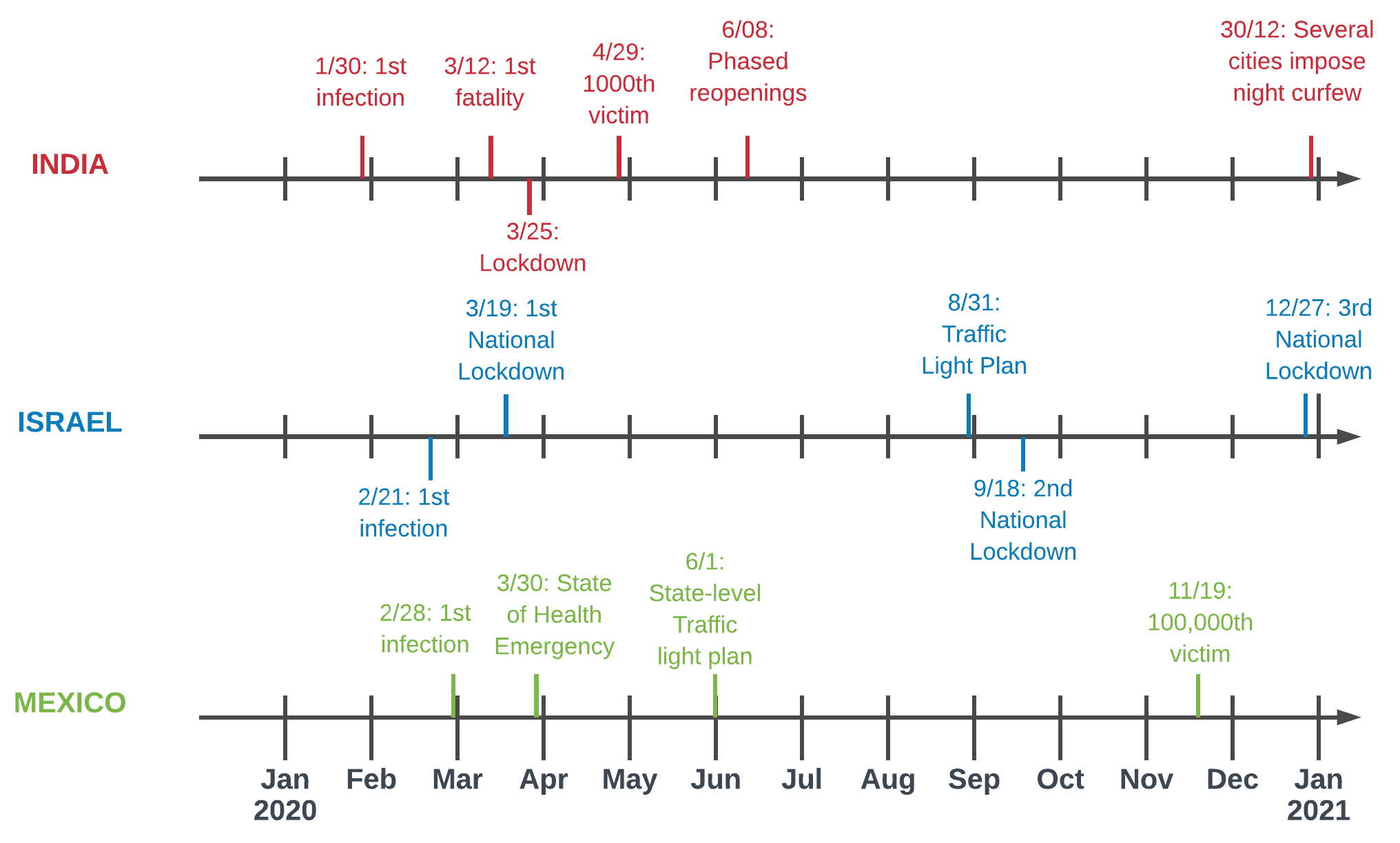}
    \caption{Timeline of main COVID-19-related events in India, Israel and Mexico (January 2020-December 2020)}
    \label{fig:timeline}
\end{figure}

Similarly as to India, conflicts pre-dating the pandemic were paused in the early months of COVID-19 only to later intensify.
Most notably, gang violence and drug cartel battles increased during the summer, as travel restrictions were lifted and the
drug trade reorganized itself to adapt to the new circumstances \color{black} \cite{SanchezMexicofocusescoronavirus2020} \color{black}. Mexico has also witnessed a large increase in murders
and especially femicides as COVID-19 restrictions have made women more vulnerable to domestic violence \color{black} \cite{MurrayMurderswomenMexico2020} \color{black}. Despite the pandemic, demonstrations
calling for the safety of women have continued. 
In Mexico 71$\%$ of the total population had internet access in January 2021 \color{black}\cite{DatareportalDigital2021Mexico2021}. Figure \ref{fig:timeline} synthetizes the most important events in the unfolding of the pandemic in India, Israel and Mexico.  \color{black}

\section{Materials and Methods}

\subsection{Data}
Our analyses are based on data obtained from the  
COVID-19 Disorder Tracker (CDT) initiative
of the Armed Conflict Location \& Event Data Project (ACLED) \color{black}\cite{ACLEDCOVID19DisorderTracker2020}\color{black}. ACLED is a well-known open-source data collection organization
cataloguing global crises and conflicts, created to facilitate the study of political violence and social unrest. 
Its recent CDT initiative records events that are directly related to COVID-19 and excludes disorders that
may temporally overlap with the pandemic but that are not directly related to it, such as conflicts between armed militias over a disputed territory \color{black} (see \cite{ACLEDMethodologyBriefCoronavirusRelated2020}). \color{black}
The dataset includes, for example, protests against governments in response to COVID-19 decisions, 
attacks against COVID-19 healthcare workers, or against individuals who may allegedly spread the virus to others.
Events are labeled as one of six types: violence against civilians, riots, protests, explosions, battles, and
strategic developments.  We exclude the strategic development category as it lists
contextually relevant episodes that are not political violence but that may trigger, lessen or explain them. These are typically 
tactical changes by one of the relevant actors, such a government announcing a state of emergency or the
easing of lockdown restrictions. In addition to type, the dataset records date and location of the event, the subjects involved, reported fatalities.
Data is collected from governmental institutions, news media,  humanitarian agencies, and research publications.
The database is updated every week; in this work we included events occurring between
January 3$^{\rm rd}$ to December 12$^{\rm th}$ 2020.
Within this interval, a total of 20,135 disorder events were recorded worldwide. 
When considering weekly averages we will consider Sunday as the first day of the week
and Saturday as the last. Hence, weekly averages will be conducted over the period
January 5$^{\rm th}$ to December 12$^{\rm th}$ 2020, yielding weeks 2 to 50
of the year 2020. Week 1 of 2020 is the period between December 29$^{\rm th}$ 2019 and January 4$^{\rm th}$ 2020.
Since disorder events were first recorded on January 3$^{\rm rd}$ 2020 we assume
all prior dates in week 1 carry zero events.

\begin{figure}[!h]
    \centering
    \includegraphics[scale=0.48]{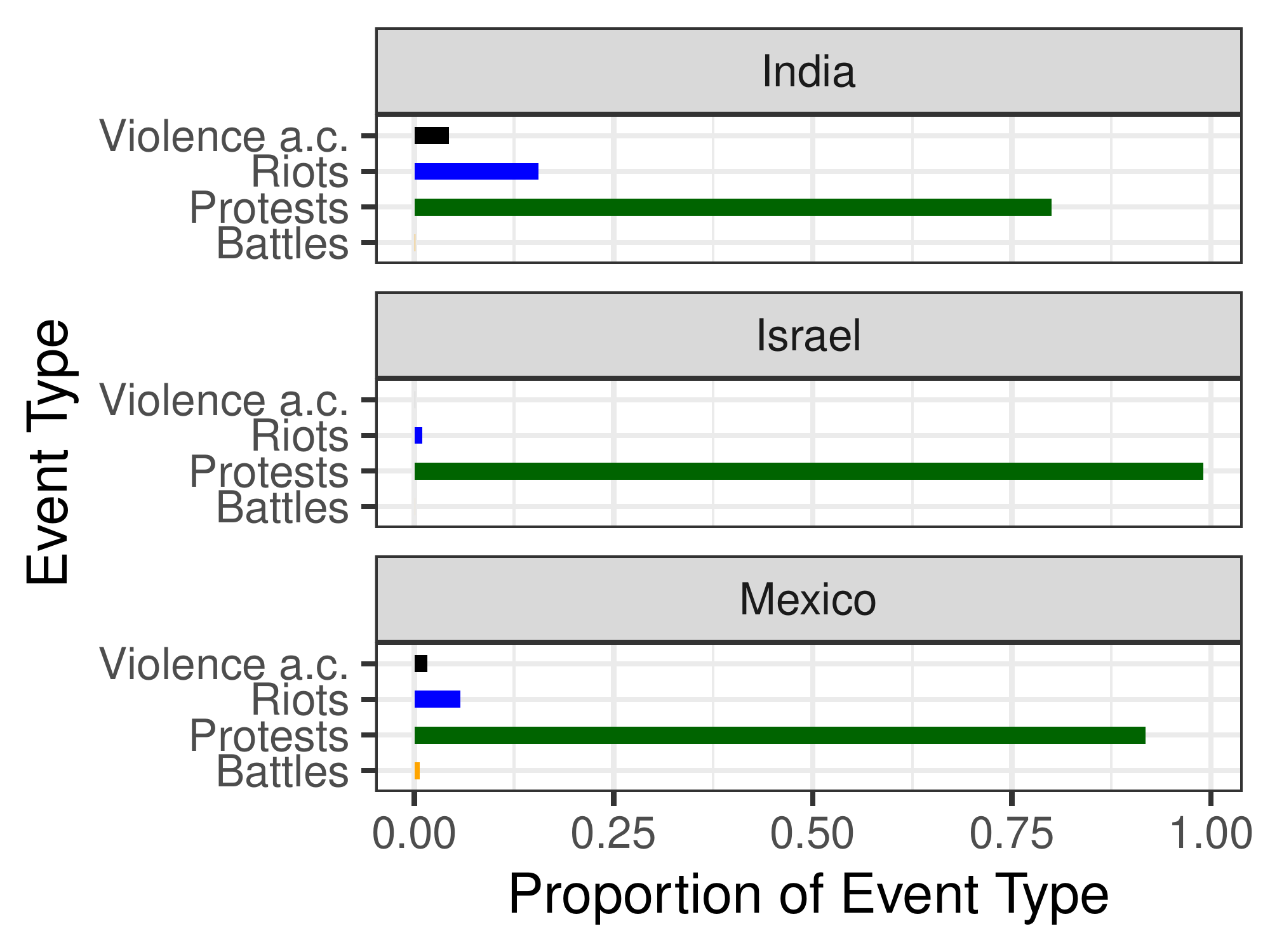}
    \caption{ Event type distribution across India, Israel and Mexico \color{black} as tallied from January 3$^{\rm rd}$ to December
12$^{\rm th}$, 2020. ``Violence a.c.'' stands for ``Violence against civilians''. \color{black} The majority of events are protests, followed
    in smaller percentage by riots, violence against civilians, and battles respectively. ACLED also lists explosions, which
    are not reported in any of the countries under investigation.}
    \label{fig:barcha}
\end{figure}

Table \ref{desc} lists the ten countries with the highest number of incidents; we focus 
on the first three: India, Israel and Mexico.  These countries alone account for almost 40\% of all the events included in the CDT dataset. 
Figure \ref{fig:barcha} shows the distribution of event type per country. As can be seen, most are protests, 
with a smaller number of riots (mostly in India and to a lesser degree in Mexico), a marginal presence of violence against civilians, 
and an almost negligible number of battles.  
No explosions were recorded in any of the countries of interest, leaving only four relevant categories
as detailed in the Supporting Information section S1.1.

\begin{table}[!t]
\centering
\footnotesize
\begin{tabular}{lcc}
\hline
\textbf{Country} & \textbf{Recorded Events} & \textbf{\%} \\\hline
\color{blue}{\textit{Israel} }& \color{blue}{\textit{3,531} }&\color{blue}{\textit{18.2}} \\
\color{blue}{\textit{India}} &\color{blue}{\textit{2,910}} & \color{blue}{\textit{15.0} }\\
\color{blue}{\textit{Mexico}} & \color{blue}{\textit{1,276}} & \color{blue}{\textit{6.6}} \\
\color{black}Argentina & 988 & 5.1 \\
Brazil & 920 & 4.7 \\
Pakistan & 588 & 3.0 \\
South Korea & 525 & 2.7 \\
Chile & 466 & 2.4 \\
Peru & 460 & 2.4 \\
Morocco & 429 & 2.2 \\
Other (128 countries) & 7,318 & 37.7\\\hline
\\
\end{tabular}
\caption{Distribution of disorder events (violence against civilians, riots, protests, battles) 
across the top 10 countries as tallied by ACLED between January 3$^{\rm rd}$ and December
12$^{\rm th}$ 2020. 
The countries considered in this study, India, Israel and Mexico are italicized, and account for almost 
$40 \%$ of the world total. At this time of writing, ACLED does not report data for the US.}
\label{desc}
\end{table}

\begin{figure}[!hbt]
    \centering
    \includegraphics[scale=0.45]{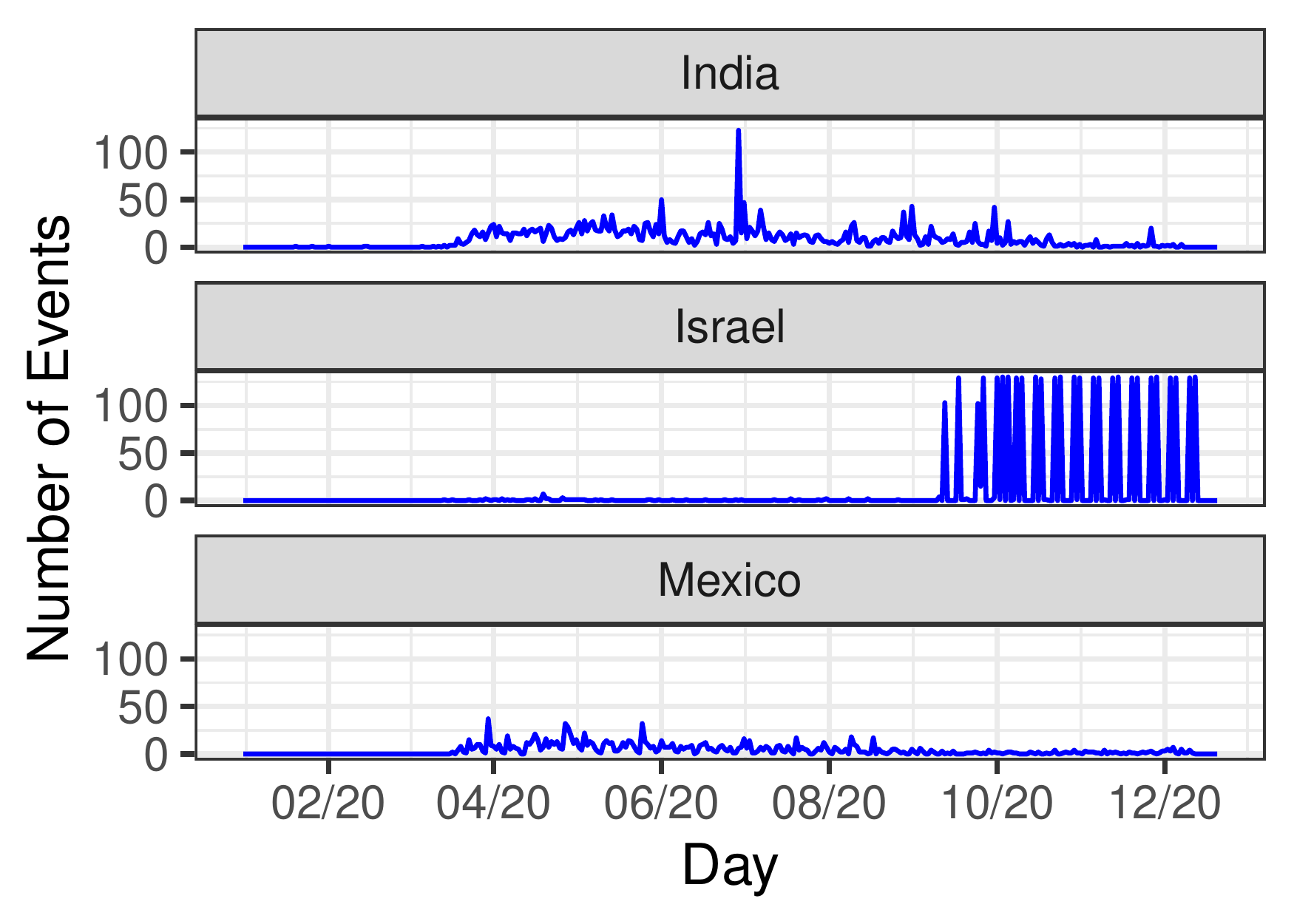}
    \caption{Time series of nationwide disorder events (protests, riots, violence against civilians, battles)
    in India, Israel and Mexico, from January 3$^{\rm rd}$  to December
    12$^{\rm th}$ 2020. Note the relatively more uniform distribution in Mexico, compared to the more structured 
    ones in India and especially in Israel.}
    \label{fig:ts}
\end{figure}

Figure \ref{fig:ts} displays the time-series of all types of events
for the countries we investigate. As can be seen, trends
differ due to the different political timelines.
In India, the first pandemic-related event was recorded in January 2020,
but was followed by only five others over the first two months of the year.  Most 
disorders in India occurred in  April (431 events, 14.8\% of the countrywide total), May (592 events, 20.3\% of the countrywide 
total), and June (481 events, 16.5\% of the countrywide total).
On June 29$^{\rm th}$ 123 protests were tallied nationwide, the largest number 
during the period of interest and corresponding
to 4.2\% of the total number of events in India. These protests were promoted by 
Congress in opposition to increasing fuel prices despite economic hardships due to the
coronavirus crisis. In Israel disorders first emerged in March 2020, however the greatest
number of demonstrations occurred in October (1,427 events, 40.4\% of the countrywide total)
and in November (1,037 events 29.4\% of the countrywide total). 
The dramatic spikes observed in this period overlap with the lockdown orders
announced at the onset of the second wave of infections; the Black Flag Movement
promoted nationwide demonstrations. In Mexico, disorders also first appeared in March 2020. However, the distribution of events here seems to be 
more uniform than in India and Israel; this may be due to Mexico never having imposed a complete, countrywide lockdown.  
The highest recorded number of disorders in the country were tallied on 
March 30$^{\rm th}$ (36 events) when health workers protested against the lack of medical supplies and equipment, and on
 April 27$^{\rm th}$  (28 events) and May 25$^{\rm th}$ (32 events) when citizens demanded better economic and financial aid.
 Overall, the most demonstrations took place in April (289 events, 22.7\% of the countrywide total)
and in May (275 events 21.6\% of the countrywide total).

\subsubsection{Spatial clustering}
Spatial-temporal concentration is a well-known signature of social phenomena, including disorder
events \cite{BaudainsSpatialPatterns20112013, DaviesmathematicalmodelLondon2013}.  
To examine the spatial distribution of disorders as listed in the CDT dataset for the three countries
of interest, we first identify subnational geographic areas where events colocalize. We do this by
using $k$-means clustering, a well-known algorithm 
which assigns each event to one of $k$ clusters by iteratively updating the centers of these clusters 
and minimizing the root-mean-square distance between the event
location and its assigned cluster center \cite{LloydLeastsquaresquantization1982a}. 
The number of clusters $k$ is a parameter for the algorithm; 
as it increases, the average mean distance of events from their assigned center
typically decreases. However, beyond a critical threshold $k^*$, the decrease rate may become negligible,
indicating that new clusters are not distinguishable from old ones.  As shown in 
Supporting Information section S1.2, applying $k$-means clustering
to the CDT data for India, Israel and Mexico, yields $k^*=4$ as this critical threshold for all three. Henceforth,
we will consider four geographical clusters per country.

\subsubsection{Hawkes process}

The temporal Hawkes process
\cite{HawkesSpectraselfexcitingmutually1971a, CoxStatisticalAnalysisSeries1966, 
BartlettSpectralAnalysisPoint1963}
models the distribution of a random time variable as a non-homogeneous Poisson process 
with a time-dependent intensity function $\lambda(t)$ defined as

\begin{equation}
\lambda(t)=\underset{h \to 0}{\mathrm{lim}}\frac{\mathbb{E}(N(t+h)-N(t))}{h}.
\end{equation}

\noindent
Here $N(t)=\sum_{i}\mathbf{1}_{t_{i}\leq t}$ is number of events up to time $t$. The index $i$ labels the events
so that the event time sequence $\{t_i \}$ defines  the point process.
Thus,  $\lambda(t)$ measures the number of events that are expected to arrive per unit time. For a homogeneous Poisson process
$\mathbb{E}(N(t)) = \lambda t$ is proportional to time and  $\lambda(t) = \lambda$ is constant.
Within non-homogeneous Hawkes processes $\lambda(t)$  is typically decomposed into 
a background intensity $\mu(t)$, often assumed to be constant $\mu(t) = \mu$, 
and an excitatory component $g(t)$ triggered at the times $t_i$ of past events so that 

\begin{equation}
    \lambda(t)=\mu + \sum_{t_i<t}g(t-t_i)
    \label{selfexcite}
    \end{equation}

We model the excitatory function $g(t)$ through an exponential decay, according to standard
protocols

\begin{equation}
    \lambda(t)=\mu + \sum_{t_i<t}\alpha\mathrm{e}^{-\beta (t-t_i)}, 
    \label{intensity_hawkes}
\end{equation}
where $\alpha$ and $\beta$ quantify the  self-excitatory degree of the process.  Here, 
$\alpha$ represents a jump factor representing the rate of increase of events immediately after
a triggering, prior event, while $\beta$ is the associated decay rate; 
$1/\beta$ is often used as a proxy for the typical lifetime of an excitation. Large
values of $\alpha$ and small values of $\beta$ imply the process is highly reactive and its effects
last longer, respectively \cite{LewisSelfexcitingpointprocess2012e}. 
 The $\alpha$ and $\beta$ parameters are learned by applying 
 maximum likelihood estimation (MLE) to Eq.\,\ref{intensity_hawkes}
 where $\{t_i \}$ is known and the $\mu, \beta,\alpha$ parameters are to be determined, 
 as described in Supporting Information Section S1.3.
From $\alpha$ and $\beta$ one can derive the branching ratio $\gamma$, an estimate of
the total number of events that are endogenously generated by a single event

\begin{equation}
    \gamma = \int_0^\infty \alpha e^{-\beta t} dt = \frac{\alpha}{\beta}.
\end{equation}

If one considers ``immigrant" events those that occur 
independently within a given generation, and the ones they trigger at the next generation 
as their ``offspring", 
$\gamma$ may also be interpreted as the total expected number of offsprings 
triggered by an immigrant event.
The so-called supercritical regime $\gamma >1$
implies that the number of offspring events is larger than the number of immigrant events that generated them,
leading to the unrealistic scenario of an infinite cascade. 
Hence, the $\gamma <1$ constraint is imposed in the maximum likelihood estimation. 
The sub-critical regime ensures that the cascade of events triggered by an original immigrant
event will eventually subside. Thus,  assuming $\gamma <1$, we can also estimate the total number of offspring events ${\cal N}_{\infty}$
generated by a single 
immigrant event. If at each generation $\gamma$ offspring events arise, the number
of events at generation $j$ is given by $\gamma^{j-1}$, where $j=1$ represents the first, single, 
immigrant event. Hence the total number of offspring events is 

\begin{equation}
\label{estimate}
    {\cal N}_{\infty}= \sum_{j=1}^{\infty} \gamma^{j-1} = \frac 1 {1 - \gamma}.
\end{equation}

Using Eq.\,\ref{estimate} we can estimate the average number of events until time $t$, due to both 
background and excitatory events

\begin{equation}
E(N(t)) \approx \frac{\mu t}{1- \gamma}, 
\end{equation}
from which the average expected number of events per unit time can be evaluated

\begin{equation}
 \lim_{t \to \infty} \frac{E(N(t))}{t} \approx 
 \frac{\mu}{1- \gamma} = \mu \left(1 + \frac{\gamma}{1 - \gamma} \right).
 \label{unittime}
\end{equation}
 
The last equality in Eq.\,\ref{unittime} implies that $\gamma <1$ also represents the percent 
of events per unit time that are endogenously generated.
We apply Eq.\,\ref{selfexcite} to our data, both as a 
baseline Poisson process (setting the self-excitability function $g=0$)
so that $\lambda(t) = \lambda =\mu$ and making the process Markovian,
and as a Hawkes process ($g \neq 0$, and as in Eq.\,\ref{intensity_hawkes}). 
To compare results, we determine the Akaike Information Criterion (AIC) 
values of the two models defined as 

\begin{equation}
\mathrm{AIC}=2 \kappa -2\mathrm{log}L
\end{equation}
where $\kappa$ is the total number of parameters used $(\kappa = 1$ for a Poisson process 
and $\kappa = 3$ for a Hawkes process) 
and $\mathrm{log} L $ is the MLE of the model
as described in Supporting Information section S1.3. By construction, the model with the lowest AIC value is the one that best fits the data. 

We also employ residual analysis to validate the choice of the exponential function $g$ in modeling 
the self-excitability of the process \cite{OgataStatisticalModelsEarthquake1988b}
Given the intensity function $\lambda(t)$ of a Hawkes
process and the set of event times $\left \{ t_i \right \}$ we can derive the set of residuals $\left \{ \tau _i \right \}$ defined as

\begin{equation}
    \tau _i = \int_{0}^{t_i}\lambda (t)\mathrm{d}t
\end{equation}

It can be shown that the $\{\tau_i\}$ residuals are independent and follow a stationary Poisson process with unit rate 
\cite{OgataStatisticalModelsEarthquake1988b, PapangelouIntegrabilityExpectedIncrements1972a}. 
This implies that the inter-arrival values $Y_i=\tau_i-\tau_{i-1}$, defined for $i > 1$ and 
where $\tau_0  = 0$ is imposed, define a set of independent and exponentially distributed variables. 
Finally, it follows that the derived quantities $U_i=1-\mathrm{e}^{-Y_i}$ are also independent and uniformly distributed.
To test the goodness of fit of the Hawkes process we can thus verify whether the
$0 \leq U_i < 1$ values are indeed uniformly distributed. 
Operationally, we employ the two-tailed Kolmogorov-Smirnov (KS) test, a common non-parametric test that
determines whether a set of given observations (in this case $\{U_i \}$) come from a known distribution (in this case the uniform
distribution between 0 and 1) \cite{ZarBiostatisticalAnalysis2010}. The test 
compares the value of the statistic $D$

\begin{eqnarray}
\label{KS}
D = {\rm max} \left[
\left({\rm max_i}  \left \vert U_i - \frac{i-1}{{\cal U}} \right \vert \right),
\left({\rm max_i}
\left \vert U_i - \frac{i}{{\cal U}} \right \vert
\right)
\right],
\end{eqnarray}
to a given critical value $D_{\rm c}$. In Eq.\,\ref{KS}, $\cal U$ is the cardinality of the
$\{ U_i \}$ set, which is the number of observations and same as the cardinality of the $\{Y_i \}$ and $\{\tau_i\}$ sets. 
If $D > D_{\rm c}$, then the hypothesis that the $\{U_i\}$ values follow a uniform
distribution, and hence that the $\{t_i\}$ values define a Hawkes-like point process,
can be rejected at the $\alpha$ level of significance.
We apply the KS test to our data at the $95 \%$ 
$(D_\alpha = 1.36/\sqrt{\cal U}$, $\alpha =0.05)$ and at the $99 \%$ 
$(D_\alpha = 1.63/\sqrt{\cal U})$, $\alpha =0.01$) confidence levels.
Other tabulated values for $D_{\rm c}$ and different confidence levels
can be used  \cite{OConnorPracticalreliabilityengineering2012}.

\section{Results}

\subsection{India}

\begin{figure}[hbt!]
    \centering
    \includegraphics[scale=0.4]{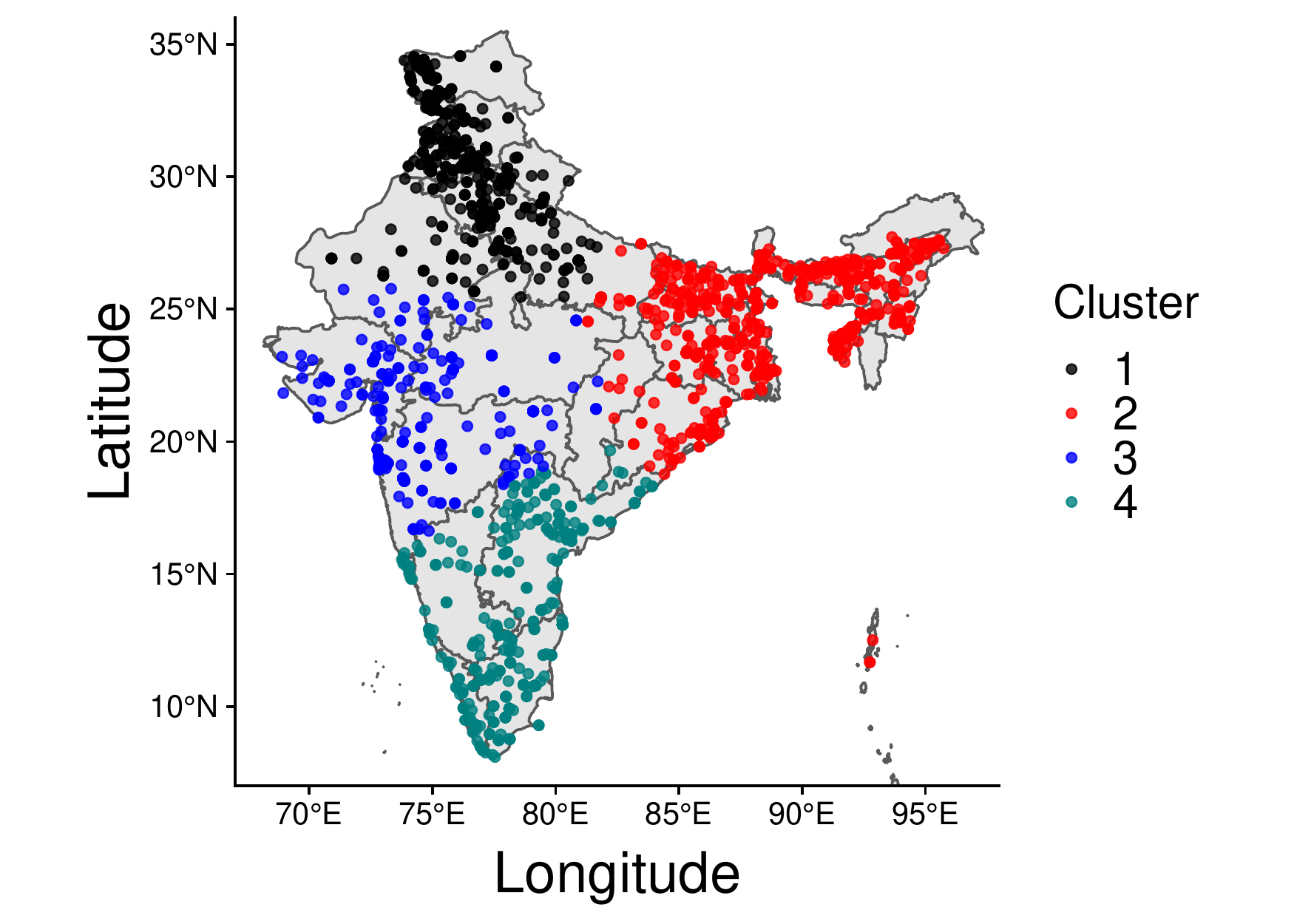}
    \caption{COVID-19 disorder events in India. The four detected clusters, C1-C4,  host approximately equal numbers of residents, with a maximum of about $27 \%$ of the total population in C1 
    and a minimum of about $18 \%$ in C2. Clusters however are very 
heterogeneous in terms of population density and territorial extent. Two of the most densely inhabited
states in the area, Uttar Pradesh and Bihar, are located in C1 and C2, respectively. \color{black} This map has been generated via $\mathrm{rnaturalearth}$ in $\mathrm{R}$, a package built using Natural Earth map data. \color{black}}
    \label{fig:india_clust}
\end{figure}

The four spatial clusters we identified in India are visualized in Figure \ref{fig:india_clust}; 
for the most part they follow geographical and/or topographical divisions within the country. 
Cluster 1 (C1), which includes the northern states of Himachal Pradesh, Rajasthan, Haryana, Uttar Pradesh, Uttarkhand, and the 
Union territories of Jammu and Kashmir and Ladakh, accounts for a total of 913 events. Cluster 2 (C2) \color{black}is associated with the highest number of events overall\color{black}, 946, and covers the eastern states of Arunachal Pradesh, Assam, Bihar, Jharkhand, and Odisha. The third cluster (C3) groups 436 in the central and western states 
of Maharashtra, Gujarat, and Madhya Pradesh. Finally, Cluster 4 (C4) accounts for a total of 568 events in the southern states, 
including Kerala, Tamil Nadu, Andhra Pradesh, Karnataka, and Telangana.

\begin{figure}[!hbt]
  \centering
  \includegraphics[width=.5\linewidth]{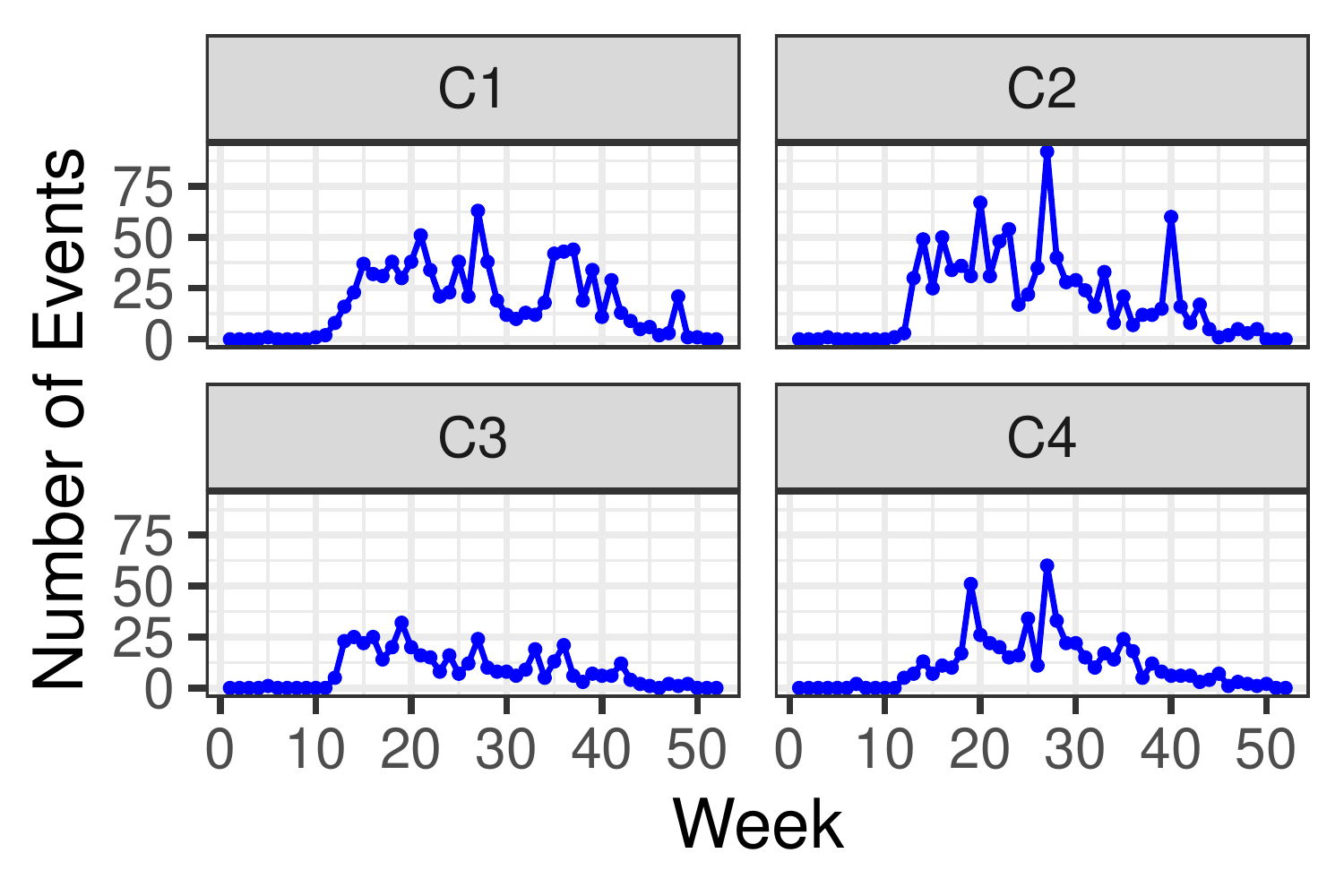}    
  \caption{Weekly time series of disorder events $\{n_j\}$ visualized by cluster, in India. Weeks are marked 
  from week $j=1$ (December 29$^{\rm th}$ 2019 to January 4$^{\rm th}$ 2020) to  
 week $j=50$ (December 6$^{\rm th}$ to December 12$^{\rm th}$ 2020). }
    \label{fig:indiats1}
  \end{figure}

Figure \ref{fig:indiats1} displays the number of events $n_j$ occurring during week $j$ in each of the C1-C4 clusters. 
For all clusters we observe the first spiking of events around week $j=13$ and $j=14$ (between March 22$^{\rm nd}$ and April 4$^{\rm th}$ 2020) followed
by a second period of intensifying activity around week $j=27$ (between June 28$^{\rm th}$ and July 4$^{\rm th}$), although the relative
magnitude of events are cluster-dependent \color{black}. Cluster-wise event distributions for India are also visualized in box plots in Supporting Information section S1.4.1. \color{black} 

\begin{figure}[!hbt]
  \centering
\includegraphics[scale=.4]{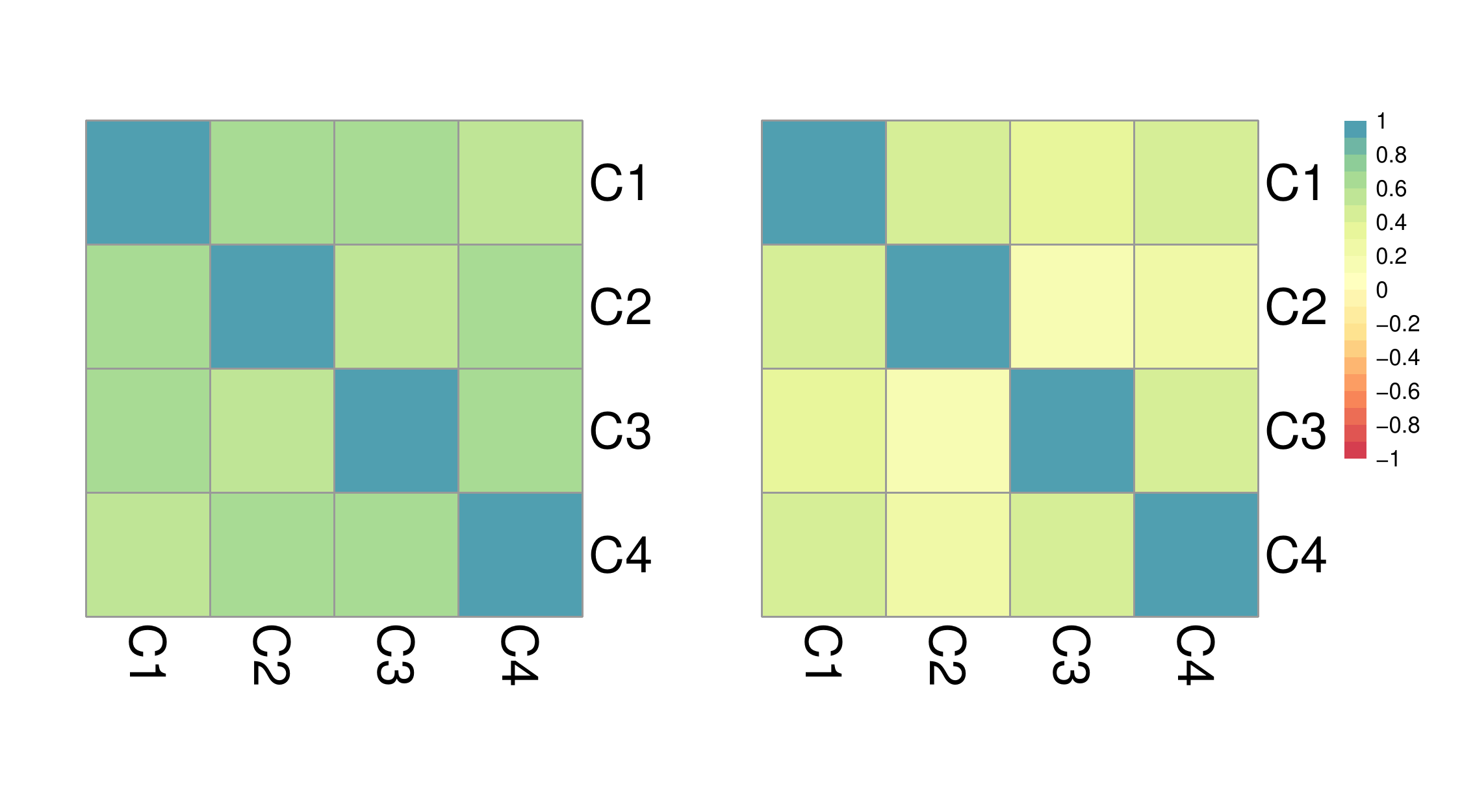} 
\vspace{-1cm}
  \caption{Cluster dynamics in India: (left) Pearson's correlation of weekly events 
  $n_{j}$ across pairs of clusters. (right) Pearson's correlation of differentiated weekly events 
 $\Delta n_j = n_{j}-n_{j-1}$.  The color scale is restricted to positive values as as no negative relationships are found. 
 The left panel reveals a substantial level of correlation, however, when first-order differences are 
 considered all coefficients decrease, implying that the rate of change in the occurrence of 
 events is less correlated.}
 \vspace{0.2cm}
\label{fig:indiats3}
  \end{figure}

The left panel of Fig.\,\ref{fig:indiats3} displays
Pearson's correlation coefficient $r$ for the weekly number of events $\{n_j \}$ between pairs of clusters.
This quantity ranges from $r=0.595$ between C2 and C3 to  
$r=0.724$ between C1 and C3, showing relative synchrony among  clusters.
Starting from $\{n_j\}$ we can also construct the first-order difference sequence  $\{\Delta n_j\}$ where
$\Delta n_j = n_{j}-n_{j-1}$, and investigate how differences in weekly counts correlate between clusters.
The right panel of Fig.\,\ref{fig:indiats3} shows that when using
$\{\Delta n_j \}$ the positive association between clusters lowers significantly compared to when $\{n_j\}$ is used, 
but remains significant in some clusters indicating partial national synchrony.  Specifically, C3 and C4 carry the highest 
correlation value, $r=0.512$, whereas the weakest relationship is between C1 and C2, for which $r= 0.124$. 
Noteworthy is the relationship between C1 and C3, which has the highest $r$ relative to $\{n_j\}$ counts 
and the second-lowest $r$ relative to $\{\Delta n_j\}$ counts. Numerical values are listed in 
Supporting Information subsection S1.5.1.

Finally, in Table \ref{india_h} we list statistical outcomes pertaining to the India disorders, where we tally events daily.
Here, the Hawkes process with an exponential excitatory term $g$ always leads to lower AIC values
compared to the baseline Poisson process, indicating a considerable degree of self-excitability and temporal dependence 
in the data. All cases considered (national and subnational Hawkes processes) passed the KS test, 
as can be verified by the $D$ values being lower than the critical KS value in Table \ref{india_h}.
We tally 2,910 countrywide disorders,  which if treated as a Hawkes process on the national scale, 
lead to a background rate of $\mu = 0.5$  events per day with 
self-excitatory events arising at a rate of $\alpha = 2.07$ events per day, spanning over $1/\beta = 0.46$ days.
These values lead to a branching ratio $\gamma = \alpha/\beta = 0. 95$, 
revealing that a large percentage of disorder events are due to feedback mechanisms. 

\begin{table}[!htb]
\footnotesize
\centering
\begin{tabular}{c|l|ccccc}
\hline
\multicolumn{2}{c|}{\bf Cluster} & \textbf{\begin{tabular}[c]{@{}c@{}} \,India \, \\ (all)\end{tabular}} & \textbf{\begin{tabular}[c]{@{}c@{}} \,India \, \\ (C1)\end{tabular}} & \textbf{\begin{tabular}[c]{@{}c@{}} \, India \, \\ (C2)\end{tabular}} & \textbf{\begin{tabular}[c]{@{}c@{}} \,India \, \\ (C3)\end{tabular}} & \textbf{\begin{tabular}[c]{@{}c@{}}
\, India \, \\ (C4)\end{tabular}} \\ \hline
 & Number of events & 2,910 & 913 & 993 & 436 & 568 \\ 
 \cline{2-7} 
 & $\mu$ & { 0.497} & { 0.611} & { 0.368} & { 0.173} & { 0.322} \\
  & $\alpha$ & { 2.073} & { 1.529} & { 1.400} & { 0.569} & { 0.712} \\ 
  & $\beta$ & { 2.192} & { 1.928} & { 1.586} & { 0.646} & { 0.854} \\
\cline{2-7} 
 & $\gamma$ & { 0.946} & { 0.793} & { 0.882} & { 0.881} & { 0.834} \\ & $\mu/(1-\gamma)$ & { 9.15} & { 2.95} & { 3.13} & { 1.44} & { 1.93} \\
 & Hawkes AIC & { -9298} & { -768} & { -1178} & { 286} & { 23} \\
 & Poisson AIC & { -6983} & { -144} & { -261} & { 579} & { 403} \\ \cline{2-7} 
 & KS Stat, $D$ & { 0.049} & { 0.051} & { 0.084} & { 0.125} & { 0.063} \\
 & KS Crit 95\%, $D^{95}_{\rm c}$ & { 0.112} & { 0.098} & { 0.127} & { 0.205} & { 0.142} \\
 & KS Crit 99\%, $D^{99}_{\rm c}$ & { 0.135} & { 0.118} & { 0.152} & { 0.245} & { 0.170} \\
\hline
\end{tabular}
\vspace{0.5cm}
\caption{Statistical outcomes of the Hawkes process applied to data from India.
The Hawkes process outperforms the baseline Poisson process both nationwide and in each 
cluster, since the Hawkes AIC is always less than the Poisson AIC. The Hawkes process 
passes the KS test at the 95\% significance level in all cases, with $D < D^{95}_{\rm c}$.}
\label{india_h}
\end{table}

On the more local level, C1 and C2, the clusters with the largest number of events,
display similar self-excitatory trends compared to C3 and C4. As can be seen
from Table \ref{india_h} values of $\alpha$ are larger in C1 and C2 ($\alpha = 1.52$ and 1.40 events per day, respectively) 
than in C3 and C4 ($\alpha = 0.57$ and $0.71$ events per day, respectively) 
however the associated timescales $1/\beta$ are less than one day in C1 and C2
 ($1/\beta =0.5$ and $0.6$ days, respectively), smaller compared to those observed for
C3 and C4 ($1/\beta =1.54$ and $1.2$ days, respectively). These results suggest that C1 and C2 are marked by
 more intense, yet more \color{black} quickly \color{black} damped feedback activity than C3 and C4.
 The two opposite trends lead to relatively uniform branching ratios across clusters with
 $\gamma$ ranging  from $\gamma=0.88$ (C2) to $\gamma = 0.79$ (C1).  
Of particular interest is C3, the most sparse cluster, as can be seen in Figure \ref{fig:india_clust}). This cluster also
carries the least number of events (436) and displays the lowest reactivity 
yet, it displays the longest timescale, leading to a very large branching ratio
$\gamma = 0.88$.  These results suggest that although disorder events
are rarer in C3, feedback effects are very strong and echoes of disorder persist the longest.
Finally, the background intensity $\mu$ is highest in C1 ($\mu =0.61$ events per day)
and smallest in C3 ($\mu =0.17$ events per day). Overall, one may expect the emergence of
roughly 9 events per day countrywide, with a high likelihood of events being self-excited; disaggregating trends within the separate clusters
shows that most of these events are to be expected in C2 and C1, and to a lesser degree in C4 and C3. 
While C3 contributes less than others to the expected
daily disorder count, the degree of self-excitation is strong. 
Interestingly, C1 and C2 are also the clusters that according to the 2011
Census of India, host the states with the most dense population: 
Uttar Pradesh in C1 (200 million residents, and a density of 828 persons per km$^2$)
and Bihar in C3 (105 million residents, and a density of 1,102 persons per km$^2$).

  \subsection{Israel} 
\begin{figure}[!hbt]
    \centering
    \includegraphics[scale=0.4]{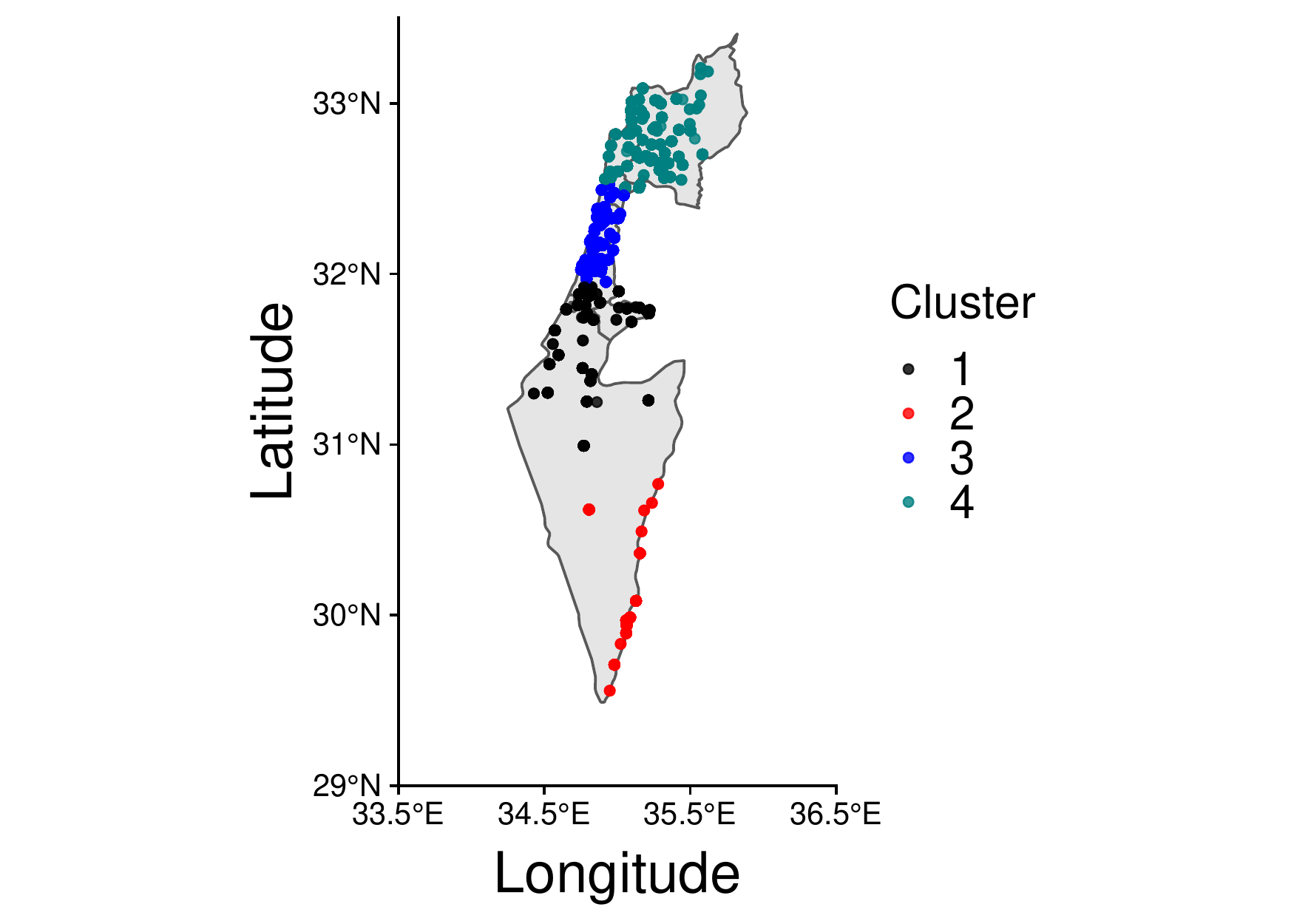}
          \caption{COVID-19 disorder events in Israel. Clusters C1, C3, C4 host the most
          densely populated areas located around the cities of Haifa, Tel Aviv and Jerusalem,  respectively,
          Events in cluster C2, the least dense region, are the most sparse and
          emerge mostly at the border with Jordan. \color{black} This map has been generated via $\mathrm{rnaturalearth}$ in $\mathrm{R}$, a package built using Natural Earth map data.\color{black}}
    \label{fig:isra_clust}
\end{figure}

  We apply similar procedures for disorder events in Israel, leading to the four clusters
displayed in Fig.\,\ref{fig:isra_clust}.  
Cluster 1 (C1) groups 792 events occurring in the greater Jerusalem area and in proximity of the Gaza strip; the second cluster (C2) is located in the southern part of the country, at the border with Jordan and only counts 220 events. The third cluster (C3)
is centered around Tel Aviv, on the western coast and contains 1073 events. 
Finally, cluster C4 is located in the Haifa region, in the northern part of the country,
and contains the most number of events, 1446.

\vspace{0.5cm}

\begin{figure}[!h]
  \centering
  \includegraphics[width=.5\linewidth]{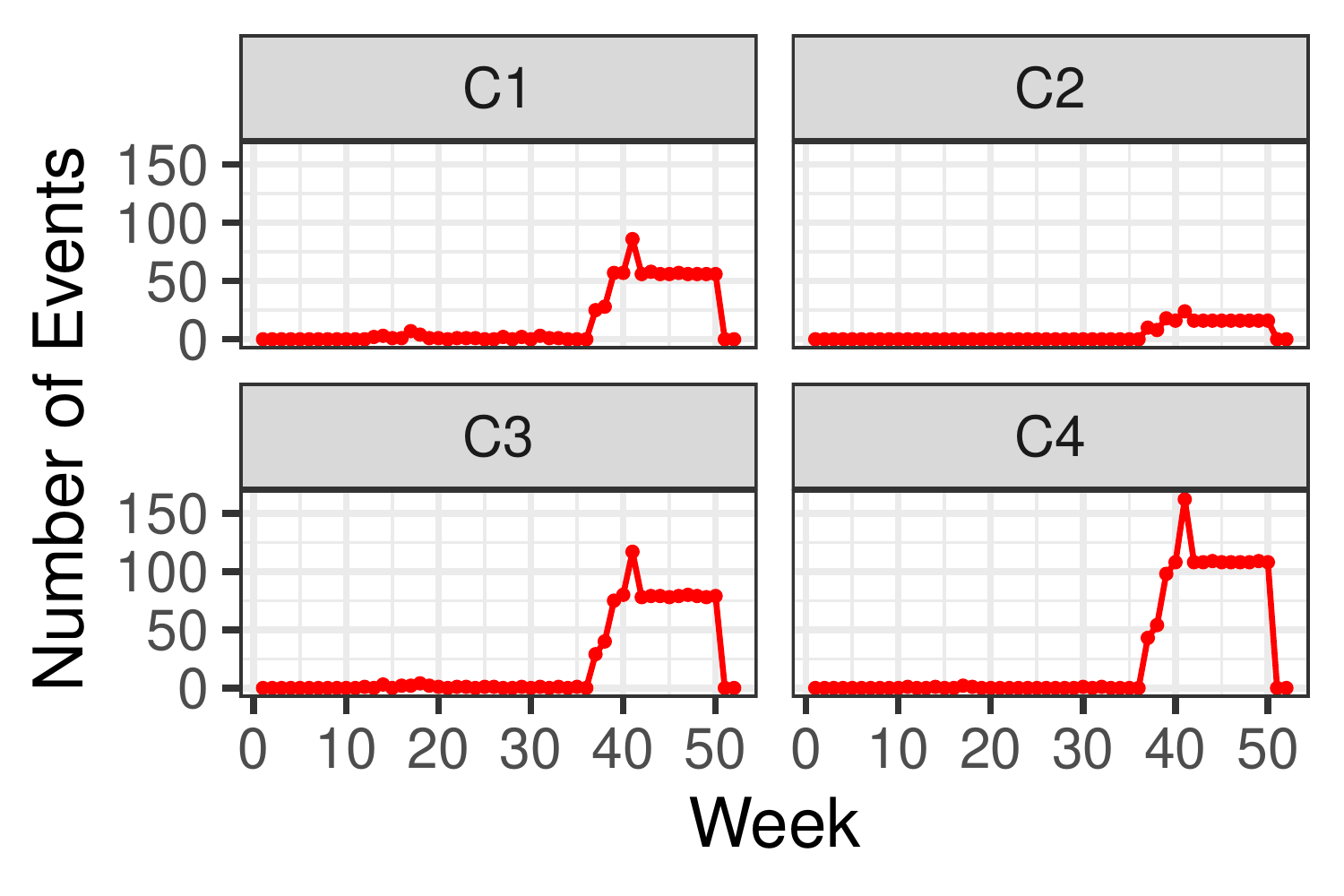}    
  \caption{Weekly time series of disorder events $\{n_j\}$ visualized by cluster, in Israel. Weeks are marked 
  from week $j=1$ (December 29$^{\rm th}$ 2019 to January 4$^{\rm th}$ 2020) to  
 week $j=50$ (December 6$^{\rm th}$ to December 12$^{\rm th}$ 2020). }
    \label{fig:israts1}
  \end{figure}

Figure \ref{fig:israts1} shows the temporal dynamics across the four clusters in Israel. 
From a temporal distribution standpoint, the overall picture is instead sensibly different from what observed in India. Besides scattered events
recorded in early 2020, most weekly events $n_j$ are concentrated
between week $j=37$ (September 6$^{\rm th}$ to September 12$^{\rm th}$)
and week $j=50$ (December 6$^{\rm th}$ to December 12$^{\rm th}$),
concurrent with nationwide protests organized by the Black Flag Movement 
against the alleged corruption of Prime Minister Nethanyahu and his failures in managing the 
pandemic \color{black} Cluster-wise event distributions for Israel are also visualized in box plots in Supporting Information section S1.4.2 \color{black} . 

\begin{figure}[!h]
  \centering
 \includegraphics[scale=0.5]{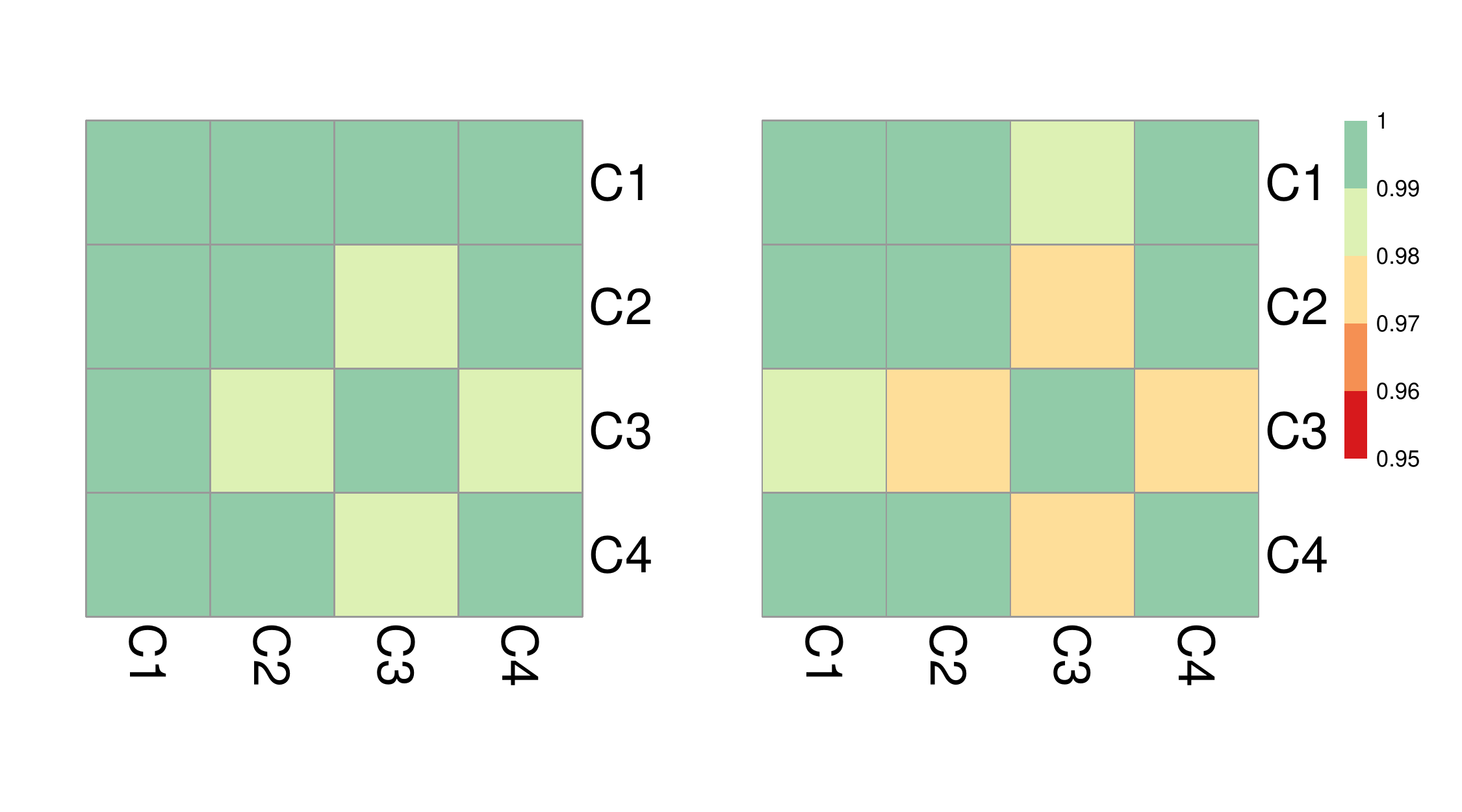}
\caption{Cluster dynamics in Israel: (left) Pearson's correlation of weekly 
events $n_j$ across pairs of clusters. (right) Pearson's correlation of differentiated weekly 
events $\Delta n_j = n_{j}-n_{j-1}$ across pairs of clusters.
 The left panel shows almost perfect correlation between weekly-based streams of events for all cluster pairs.
The synchrony remains almost perfect when considering first-order differences in the right. The nationwide correlation that is much more
visible than in India or Mexico may be due
to Israel's more compact geographical extension, linguistic unity, tighter virtual connectivity,
and/or due to the nationwide engagement
of the Black Flag movement.  \color{black} Note that correlation coefficients between clusters in Israel
are very large (r $\geq$ 0.95 in all cases) compared to those computed for India (and Mexico).
Thus, if we kept the same scale as in Fig. \ref{fig:indiats3} and Fig. \ref{fig:mexicots3} (-1 $\leq$ r $\leq$ 1)
the correlation plots for Israel would be colored uniformly. Instead, for a more nuanced view we use instead
a more restricted scale (0.95 $\leq$ r $\leq$ 1). \color{black} }
\vspace{0.2cm}
\label{fig:israts3}
  \end{figure}

The Pearson's correlation coefficients $r$ comparing weekly $\{n_j\}$ events between pairs of clusters are shown in Fig.\,\ref{fig:israts3} 
and reveal strong homogeneous, positive synchrony
Values range between $r=0.995$ (between C2 and C3, and  C2 and C4)  
and $r=0.999$ (between C3 and C4). Similarly strong, positive relationships persist when considering the first order difference sequence 
$\{\Delta n_j\}$. 
This finding underscores that even when considering increasing or decreasing trends, 
clusters are tightly aligned, indicating nation-wide synchrony. \color{black} The fact that \color{black} this feature emerges with so much clarity in Israel compared
to India may be a consequence of its much smaller territorial extent,  a more linguistically homogenous
population, and/or larger access to broadband internet, but also the result of the Black Flag Movement's nation-wide campaigns.
Numerical values are listed in Supporting Information subsection S1.5.2.

Table \ref{israel_h} lists all statistical quantities derived from fitting the Hawkes process to the tabulated disorder events in Israel.
As observed for India, the Hawkes process always outperforms the baseline Poisson process in terms of AIC values.
However, the $D$ statistic derived from the KS
analysis is higher than the critical $D_{\rm \alpha}$ both 
at the 95$\%$ and at the 99$\%$ confidence levels, 
indicating that more appropriate time-dependent 
point processes should be used to describe the data. For example, the 
extreme clustering that characterizes events in C4 might be better represented
by more rapidly decaying forms than the exponential decay $g(t)$.
From Table \ref{israel_h} we also observe that the 3,531 
countrywide events can be described by a Hawkes process with a decaying exponential function $g$
where $\alpha = 23.53$ events per day, lasting $1/ \beta = 0.45$ days. These values 
correspond to a branching ratio $\gamma = 0.97$ and indicate that disorders cause strong
feedback of short duration. This is confirmed by Fig.\,\ref{fig:ts} where we observe a very high concentration of 
events occurring over a limited timeframe. Clusters C1, C3 and C4 display relatively similar trends:
the background rate $\mu$ varies between $\mu = 0.151$ (C4) and $\mu = 0.265$ (C1) events per day, 
whereas $\gamma$ ranges from $\gamma = 0.92$ (C1) to $\gamma= 0.97$  (C4), indicating relatively low background rates
but sustained feedback. Conversely,  C2 reports the largest background rate among all clusters
with $\mu = 0.59$ events per day, the lowest reactivity with
$\alpha = 4.82$ events per day and the largest lifetime $1/\beta = 0.15$ days, which combine
to yield the lowest branching ratio $\gamma = 0.76$ of all clusters. 
On average, the total number of disorders expected in Israel is 
$\mu/(1- \gamma) = 13.8$ events per day, arising in descending
order in C3, C1 and C2. However, the contribution of 
C4 cannot be determined since the Hawkes process may not be 
the most adequate representation of the local point process distribution
in this cluster.

\begin{table}[!t]
\centering
\footnotesize
\begin{tabular}{c|l|ccccc}
\hline
\multicolumn{2}{c|}{\bf Cluster} & \textbf{\begin{tabular}[c]{@{}c@{}}  \, Israel \, \\ (all)\end{tabular}} & \textbf{\begin{tabular}[c]{@{}c@{}}  \, Israel \, \\ (C1)\end{tabular}} & \textbf{\begin{tabular}[c]{@{}c@{}} \, Israel \, \\ (C2)\end{tabular}} & \textbf{\begin{tabular}[c]{@{}c@{}}  \, Israel \, \\ (C3)\end{tabular}} & \textbf{\begin{tabular}[c]{@{}c@{}}
 \, Israel \, \\ (C4)\end{tabular}} \\ \hline
 & Number of events & 3,531 & 792 & 220 & 1,073 & 1,446 \\ \cline{2-7} 
 & $\mu$ & {0.390} &  0.265 &  0.590 &  0.236 &  0.151 \\
 & $\alpha$ &  23.528   & 9.894 &  4.822 &  12.386 &  15.306 \\
 & $\beta$ &  24.212 &  10.812 &  6.310 &  13.118 &  15.717 \\ \cline{2-7} 
 & $\gamma$ &  0.972 &  0.915 &  0.764 &  0.944 &  0.974 \\
&  $\mu/(1-\gamma)$ &  13.79 &  3.12 &  2.50 &  4.21 &  5.79 \\
 & Hawkes AIC &  -25047 &  -2655 &  -145 &  -4588 &  -7649$\dagger$ \\
 & Poisson AIC &  10988&  -149&  58 &  -805 &  -1917 \\ \cline{2-7} 
 & KS Stat, $D$ &  0.098 &  0.157 &  0.134 &  0.156 &  {\color{red}0.274} \\
 & KS Crit 95\%, $D^{95}_{\rm c}$ &  0.127 &  0.160 &  0.190 &  0.168 &  0.212 \\
 & KS Crit 99\%, $D^{99}_{\rm c}$ &  0.153 &  0.192 &  0.228 &  0.202 &  0.254 \\
\hline
\end{tabular}
\vspace{0.5cm}
\caption{Statistical outcomes of the Hawkes process applied to data from Israel.
The Hawkes process outperforms the baseline Poisson process both nationwide and in each 
cluster, since the Hawkes AIC is always less than the Poisson AIC. The Hawkes process 
passes the KS test at the 95\% significance level in all cases except for C4, where $D > D^{99}_{\rm c}$, indicating
that the hypothesis that the data can be fit to a Hawkes process with a decaying exponential should not be accepted.
Given the nature of the data, a more steeply decaying function than the decaying exponential should be used in C4.}
\label{israel_h}
\end{table}

\subsection{Mexico}

\begin{figure}[h!]
    \centering
    \includegraphics[scale=0.4]{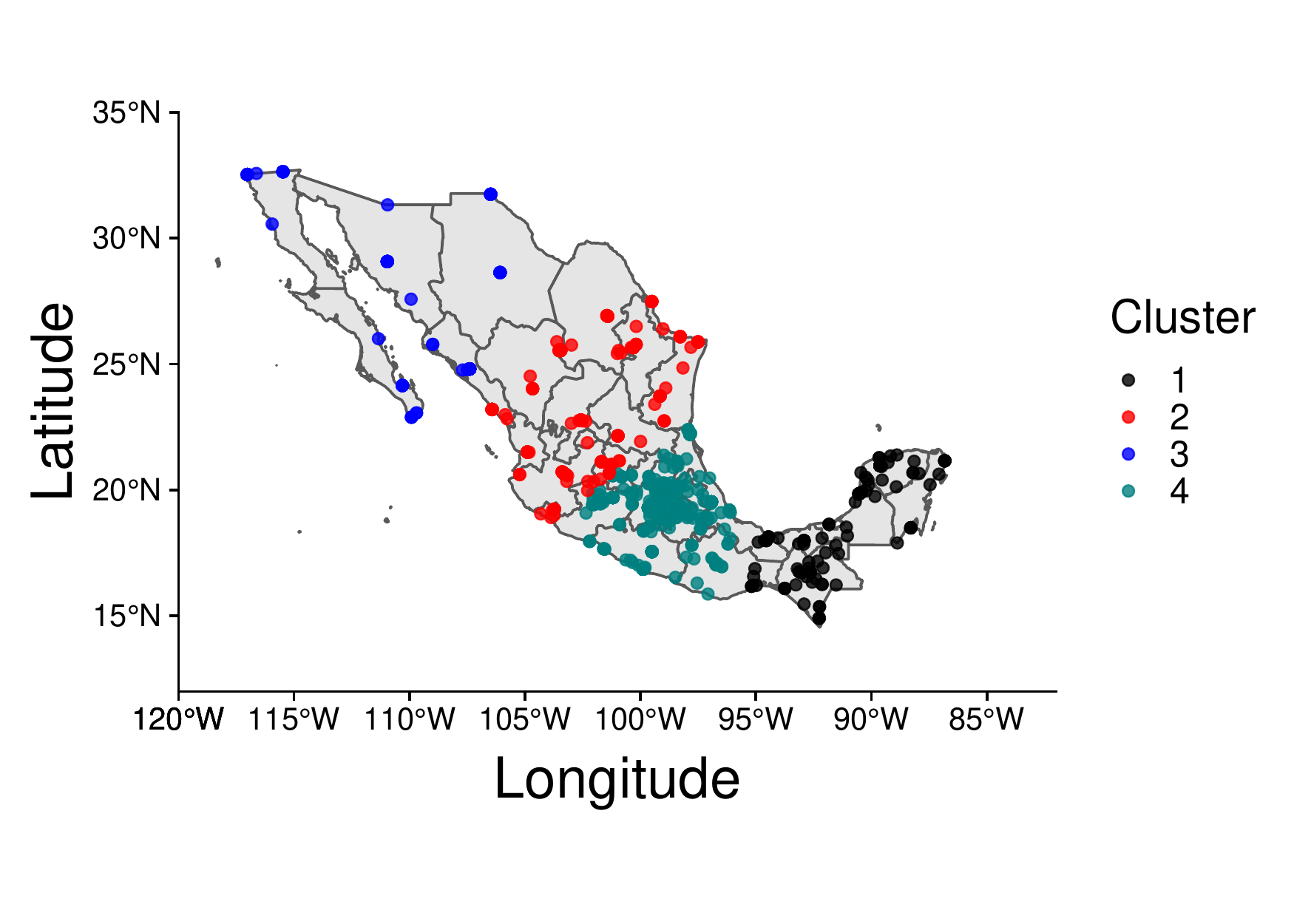}
    \caption{COVID-19 disorder events in Mexico. Of all clusters, C4 carries the largest population
    as it includes the capital city and the state of Mexico. The two are respectively the most
    populated city and state in the country. The state of Mexico is also the most dense nationwide. This map has been generated via $\mathrm{rnaturalearth}$ in $\mathrm{R}$, a package built using Natural Earth map data.}
    \label{fig:mexico_clusters}
\end{figure}

 As for India and Israel, the clustering process yields four clusters in Mexico; they are shown in Fig.\ref{fig:mexico_clusters}. 
Two of them exhibit a sparse nature, with events distributed in a less dense manner compared to the other two.
The first cluster (C1) only contains 139 events that however are spatially concentrated. 
This cluster covers the southern part of the country, and includes the states of Chiapas, Veracruz, Campeche, Yucatan, and Quintana Roo. 
The second cluster (C2) contains a total of 270 events, mainly located in the states of Durango, Tamaulipas, Nuevo Leon, and Nayarit. 
Cluster C3 is the least populated, tallying only 101 events that are spread across the Northern states, 
particularly Baja California Sur, Baja California, Sonora, Sinaloa, and Chihuahua.
As can be seen, the vast majority of events (766) are located in C4, which accounts for 60\%
of total disorders. C4 contains
Mexico City, the country's capital, as well as the states of Oaxaca, Puebla, Queretaro, Guanajuato, Michoacan, and 
Mexico state. 

\begin{figure}[!h]
  \centering
  \includegraphics[width=.5\linewidth]{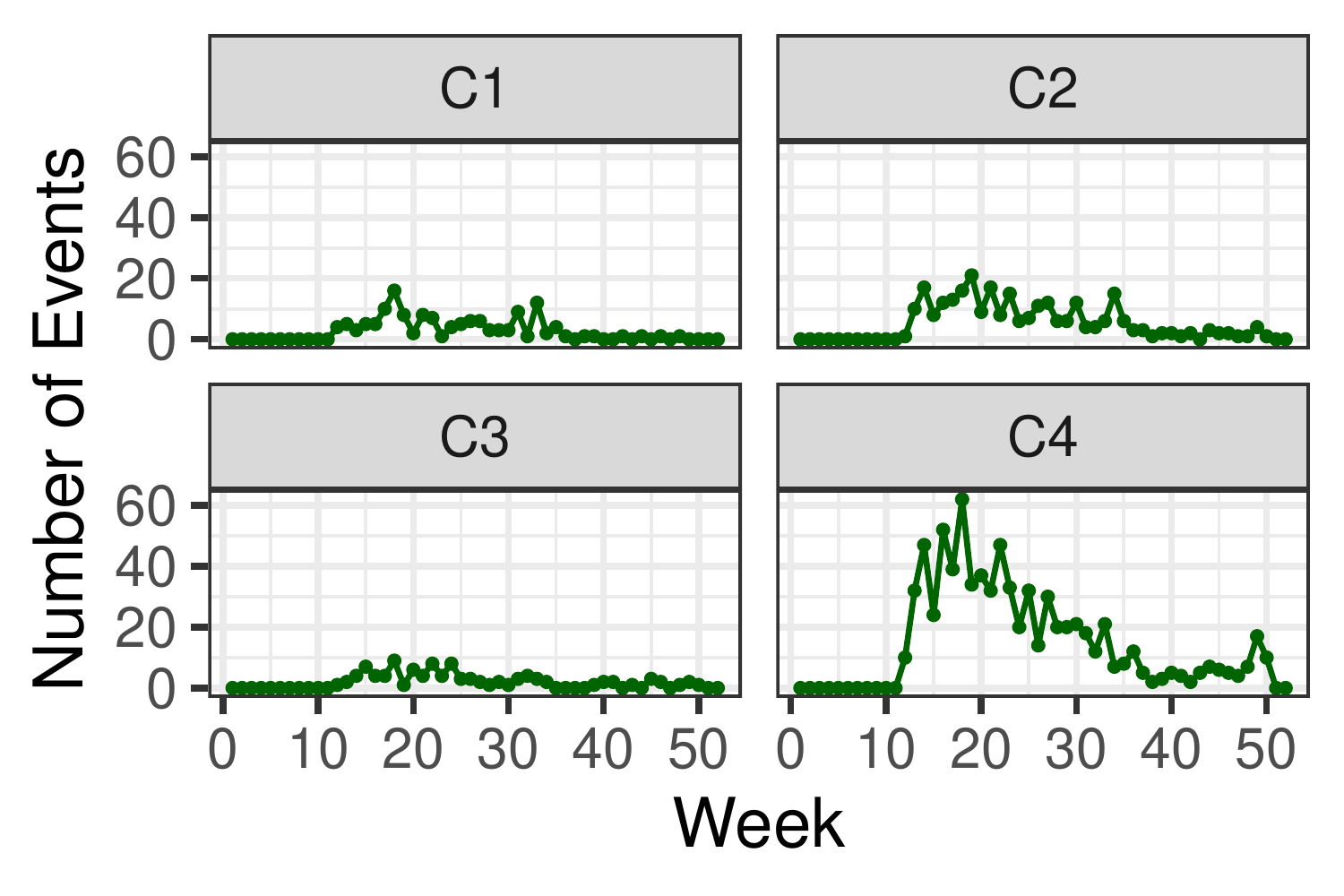}    
 \caption{Weekly time series of disorder events $\{n_j\}$ visualized by cluster, in Mexico. Weeks are marked 
  from week $j=1$ (December 29$^{\rm th}$ 2019 to January 4$^{\rm th}$ 2020) to  
 week $j=50$ (December 6$^{\rm th}$ to December 12$^{\rm th}$ 2020). }
    \label{fig:mexicots1}
  \end{figure}

Figure \ref{fig:mexicots1} displays cluster trends at the weekly level. Contrary to what observed for India
and Israel, disorder events have mostly persisted throughout the COVID-19 crisis: 
in Mexico  there have been no major interruptions of protests, riots or violence against civilians since the onset of the pandemic. 
This may be due to the government never having imposed a complete lockdown in the country.
However, the alignment in peaks of disorder activity across clusters is weak, and much less pronounced than in India or Israel. 
For instance, while C4 (the cluster that records the majority of events) shows sustained activity on 
weeks $j=14$ (March 29$^{\rm th}$ to April 4$^{\rm th}$), $j=16$
(April 12$^{\rm th}$ to April 18$^{\rm th}$), and 
$j=18$ (April 26$^{\rm th}$ to May 2$^{\rm nd}$)\color{black}, similar trends \color{black} are not detected in the other clusters. For example, 
in neighboring C2, activity spikes in week $j=19$ (May 3$^{\rm rd}$ to May 9$^{\rm th}$). \color{black} Cluster-wise event distributions for Mexico are also visualized in box plots in Supporting Information section S1.4.3. \color{black} 

\begin{figure}[hbt!]
  \centering
  \includegraphics[scale=0.5]{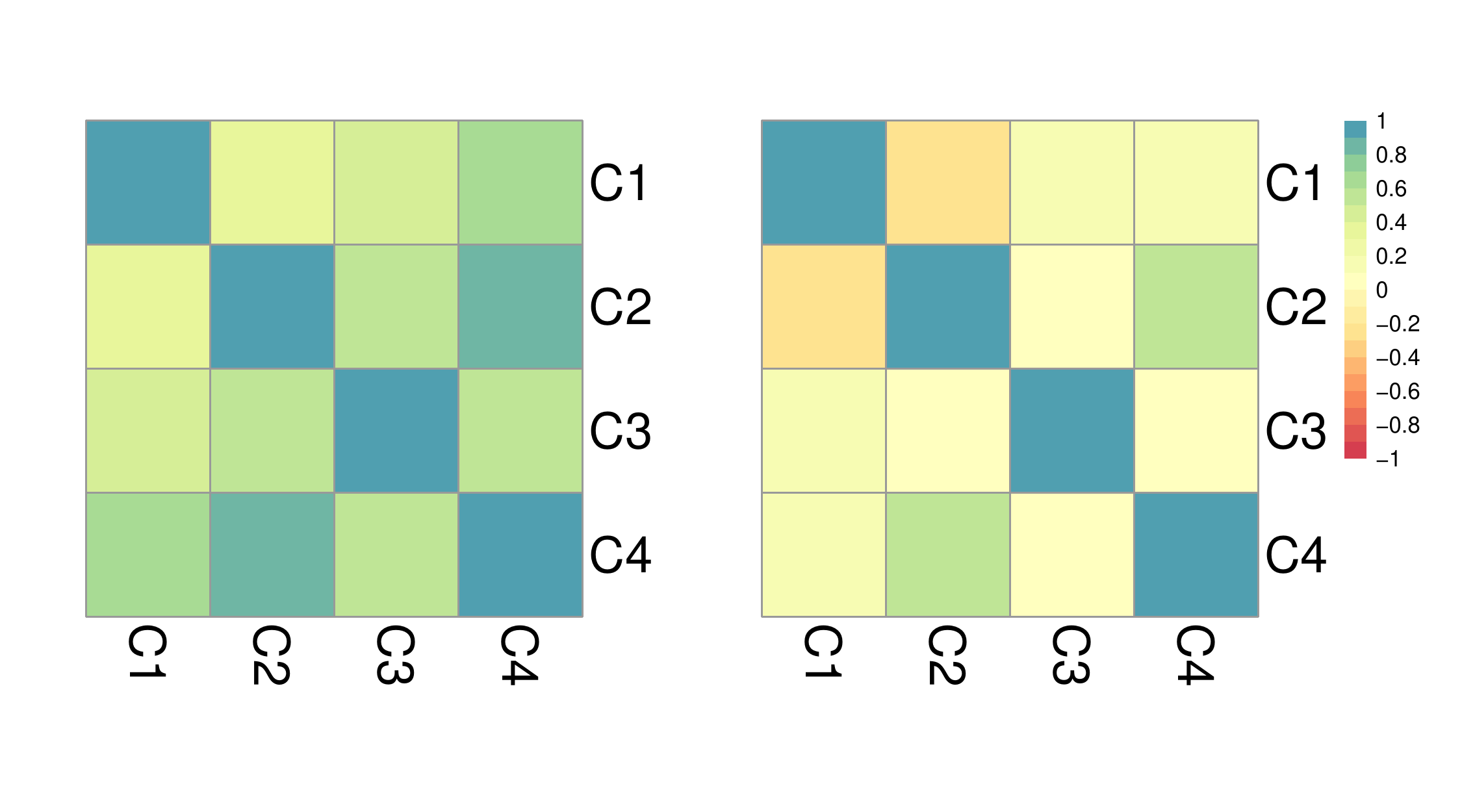}  
\label{mexcor}
\caption{Cluster dynamics in Mexico: (left) Pearson's correlation of weekly 
events $n_j$ across pairs of clusters. (right) Pearson's correlation of differentiated weekly 
events $\Delta n_j = n_{j}-n_{j-1}$. The left panel shows a high level of correlation 
between clusters; however, contrary to what observed in India and Israel,
dramatic decreases are observed when computing coefficients between first order differences 
implying a low level of synchrony in the rate of change of events. }
\label{fig:mexicots3}
\vspace{0.2cm}
  \end{figure}

The Pearson's correlation coefficients $r$ for Mexico and its four clusters are shown in Fig.\, \ref{fig:mexicots3}.
Calculations relative to the weekly $\{n_j\}$ events reveal some alignment 
between clusters, specially between C2 and C4 ($r=0.826$), and to a lesser degree between
C3 and C4 ($r=0.772$). However, after computing 
the correlation coefficient relative to the first order difference sequence $\{\Delta n_j\}$, 
these relationships become much weaker: all positive correlations between weekly counts are reduced, 
C3 and C4 display low correlation ($r=0.286$) 
and the value of the coefficients between C1 and C2 ($r=-0.092)$ and C2 and C3 ($r=-0.445$) become negative.
Numerical values are listed in 
Supporting Information subsection S1.5.3.

\vspace{0.5cm}
  
The statistical results obtained from applying the Hawkes model to Mexico are listed in Table \ref{mexico_h}.  As for India and Israel, 
the Hawkes process yields better outcomes than the baseline Poisson process in modeling disorder events related to the COVID-19 pandemic, both 
nationwide and at the cluster level.  Furthermore, all models fitted with data on Mexico passed the KS test as well, 
 testifying to the goodness of fit provided by the
specific Hawkes formalization of a self-excitability process.
We identify an underlying temporal-dependence among events even in the case of C3, the cluster with least number of events.
Countrywide, the $1,276$ events yield a Hawkes process marked by reactivity $\alpha$=2.29 events
per day, with a relatively short lifetime of $1/\beta = 0.41$ days.  The branching ratio is $\gamma = 0.79$, 
indicating a considerable amount of self-excitability, although much less than what observed in India and Israel.
Among the various clusters, the greatest reactivity is observed in C4 with $\alpha$=1.66, decreasing in other clusters
until a minimum of $\alpha$=0.10 events per day is reached in C3; C1 is also the cluster where the longest lifetime 
of self-excitatory phenomena is observed, with $1/\beta$= 6.88 days whereas the shortest is in C4
where $1/\beta= 0.44$ days. Overall, the cluster values of $\alpha, \beta$ yield
sensibly lower values of $\gamma$ compared to India and Israel, ranging
from $\gamma = 0.54$ (C2) to $\gamma = 0.73$ (C4) and confirming the nationwide trend.
The background rate $\mu$ ranges from  $\mu = 0.774$ (C4) to $\mu = 0.121$ (C3) events per day.
Combined, these results imply that nationwide one can expect a total of $\mu/(1-\gamma) =$ 4.77 events per day, 
of which the most (2.86) will occur in C4, and the least (0.39) in C3.
\newpage 

\begin{table}[!t]
\centering
\footnotesize
\begin{tabular}{c|l|ccccc}
\hline
\multicolumn{2}{c|}{\bf Cluster} & \textbf{\begin{tabular}[c]{@{}c@{}} \, Mexico \, \\ (all)\end{tabular}} & \textbf{\begin{tabular}[c]{@{}c@{}} \,Mexico \, \\ (C1)\end{tabular}} & \textbf{\begin{tabular}[c]{@{}c@{}} \, Mexico \, \\ (C2)\end{tabular}} & \textbf{\begin{tabular}[c]{@{}c@{}} \,Mexico \, \\ (C3)\end{tabular}} & \textbf{\begin{tabular}[c]{@{}c@{}}
\, Mexico \, \\ (C4)\end{tabular}} \\ \hline
 & Number of Events & 1,276 & 139 & 270 & 101 & 766 \\ \cline{2-7} 
 & $\mu$ &  0.996 &  0.217 &  0.469 &  0.121 &  0.774 \\
 & $\alpha$ &  1.931 &  0.446 &  1.086 &  0.101 &  1.660 \\ 
  & $\beta$ &  2.440 &  0.728 &  2.009 &  0.145 &  2.274 \\\cline{2-7} 
 & $\gamma$ & 0.791 & 0.612 & 0.541 & 0.693 & 0.730 \\
 &  $\mu/(1-\gamma)$ &  4.77 & 0.55 &  1.02 & 0.39 & 2.86 \\
 & Hawkes AIC &  -2201 &  382  &  413 &  377 &  -526 \\
 & Poisson AIC &  -1424&  445 &  532 &  401 &  -69\\ \cline{2-7} 
 & KS Stat, $D$ &  0.047&  0.092 &  0.076 &  0.098 &  0.063 \\
 & KS Crit 95\%, $D^{95}_{\rm c}$ & 0.082 & 0.275 & 0.123 & 0.233 & 0.092 \\
 & KS Crit 99\%, $D^{99}_{\rm c}$ & 0.098 & 0.330 & 0.147 & 0.279 & 0.111 \\
\hline
\end{tabular}
\vspace{0.5cm}
\caption{Statistical outcomes of the Hawkes process applied to data from Mexico.
The Hawkes process outperforms the baseline Poisson process both nationwide and in each 
cluster, since the Hawkes AIC is always less than the Poisson AIC. The Hawkes process 
passes the KS test at the 95\% significance level in all cases, with $D < D^{95}_{\rm c}$.}
\label{mexico_h}
\end{table}

\section{Discussion and Conclusion}

We studied COVID-19 disorder events by using a public database compiled by the CDT initiative 
promoted by ACLED, the most reliable and complete source of data on conflicts and disorder patterns worldwide. 
Specifically, we analyzed the spatio-temporal distributions and characteristics of demonstrations
in the three countries with the highest number of events {\textit {i.e.,}}
India, Israel, and Mexico between January 3$^{\rm rd}$ and December 12$^{\rm th}$ 2020.
Using the well known Hawkes point process we investigated
whether self-exciting effects could arise across events. We first considered countrywide data
and later identified distinct geographical clusters in each of the three countries,
to investigate trends on the more local level.  Our intent was to better
understand the macro- and meso-scale mechanisms that govern disorder events 
occurring in the same general context (the pandemic) but that may be ignited, shaped, acerbated
or placated by more local happenings.
Our work is in line with other empirical studies related to 
social tension,  nucleated by the 
seminal work of M. I. Midlarsky \cite{MidlarskyAnalyzingDiffusionContagion1978a} and that include  analyses of 
the 2011 London riots
\cite{BaudainsGeographicpatternsdiffusion2013, DaviesmathematicalmodelLondon2013}, and of the 2005 Paris riots
\cite{Bonnasse-GahotEpidemiologicalmodelling20052018}.

We identified four geographical clusters in each of the the countries we investigated
by employing $k$-means clustering,  These
hosted varying numbers of events, were of varying spatial extent, and mostly followed clear geographical separations.  We observed
self-excitatory effects in all countries and in almost all subnational clusters, 
and found that the time-dependent Hawkes process is always a better fit to the disorder data than
a simple Poisson process. We also performed several robustness checks, such as 
modifying the random number generator, or considering shorter time windows (see Supporting Information section S1.6); 
the number of clusters, and the applicability of the Hawkes process persisted
in all cases. These results show that temporal dependence and self-excitability at the national level are 
not the result of the superimposition of unstructured, random processes at the subnational level. 
Instead, disorder events naturally cluster already at the local level, regardless of cluster size, and country examined. 

However, while the temporal dependency between events, a hallmark of the Hawkes process, represents a common feature of protests, riots, and similar events related to the COVID-19 pandemic, important differences in the magnitude of these dependencies arise both intra- and inter-country. The three parameters that define the conditional intensity function of the Hawkes processes, {\textit {i.e.}} the background rate $\mu$, the reactivity $\alpha$, the 
\color{black}decay \color{black} rate $\beta$, as well as other derived quantities such as the
branching ratio $\gamma$ and the average expected intensity $\mathrm{E}[\lambda(t)]$, report a wide heterogeneity. 
This may be due to local infectivity trends, local decision-making, and how the pandemic and the associated measures
impacted local calendars of religious or public holidays.

The distribution of disorder events in Mexico for instance, does not display large variations
between March and October 2020; in India, although the frequency of disorders has always remained relatively large
starting in April 2020, major spikes emerged in May 2020 and between June and July 2020. 
Israel, finally, displays an even more extreme situation: a handful of events were recorded in March 2020,
but intense protesting occurred between September and October 2020.
As a result of the relatively homogenous temporal trend, Mexico is associated with the largest background intensity $\mu$; 
reactivity $\alpha$ on the other hand is largest in Israel, the country which also displays the shortest duration of an excitation
$\beta^{-1}$. In Israel, the influence of an event lasts for $\beta^{-1} =0.04$ days, ten times less than in India and Mexico 
where $\beta^{-1} = 0.46$ and $\beta^{-1} = 0.41$ days, respectively,
The more homogeneous course of events in Mexico is also manifest in the lowest branching factor, 
$\gamma = 0.791$, compared to India and Israel ($\gamma = 0.946$, $\gamma=0.972$), implying that 
the probability that an event is endogenously generated as a consequence of another event is lowest in Mexico.
Among the social and political events driving these patterns are the lack of a complete lockdown in Mexico,
large scale protests in Israel as organized by the Black Flag Movement, the more compact geography
and stronger internet connectivity in Israel compared to Mexico and India, which are also more linguistically and culturally
heterogeneous.  In all cases, the highest reactivity $\alpha$ emerges at the national scale, 
conversely, the average number of days upon which events may excite others
is always highest in subnational clusters.
\color{black} Correlating the number of events with specific cluster demographic or geographical characteristics is outside the scope of this paper. However, regions with large numbers of pandemic-related events are often, but not always, characterized by large populations and/or large population densities. One notable exception is Jammu-Kashmir, a region in India marked by a relatively large number of events that is not among the country's most populous, nor most dense areas. Similarly, while the greater Jerusalem area is among the most densely inhabited in Israel, relatively few disorder events have been recorded here. These findings underly the need to consider the intersectionality of many demographic, socioeconomic, political and/or religious factors when trying to understand why some clusters display more events than others. \color{black}

Our work comes with some limitations. First, a more systematic approach to assess temporal dependence and self-excitability on the global scale
is needed, especially considering the risk of prolonged restrictive measures during national vaccination campaigns.  Although India, Israel, and Mexico report the highest number of disorders, they represent 
a small fraction of the countries in which disorders have occurred. Evaluating whether temporal clustering is a universal characteristic would help 
advance our knowledge on human behavior under prolonged periods of duress, and under greater social and cultural diversity.
Furthermore, we did not distinguish events based upon their nature ({\it{e.g.}}, protests vs. violence against civilians), 
motives ({\it{e.g.}}, protests against restrictions imposed by the government vs. protests against the lack of supplies for health care workers;
or participant type (migrant workers vs. students). As mentioned, the vast majority of events in India, Israel and Mexico, 
have been peaceful protests, however the dataset we utilized did not allow for a clear stratification of motives or participant type.
Disentangling the distinct mechanisms that trigger self-excitability 
based on the qualitative features of the event themselves and the demands participants carry, 
would greatly add to our understanding of how social unrest unfolds. 

Notwithstanding, our work shows that analyzing disorders at the national and subnational scales 
is useful. A national focus allows \color{black}us \color{black} to understand higher-level dynamics
that transcend physical distances, especially in the current era of unprecedentedly fast information (and disinformation) spread.
Social-media and other forms of long-distance connectivity can facilitate the dissemination of
government decisions but also help large scale planning of nationwide or even international
disorders.  Understanding subnational dynamics allows for a more nuanced picture as local streams of protests may be 
ignited by more local issues. The global nature of the pandemic does not imply that its impacts are
 homogeneously distributed; to the contrary, existing socio-economic
contexts lead to heterogeneous responses to the unfolding and management of the crisis. 
Decisions taken may resonate differently in different communities, which may be more or less
concerned with the restriction of individual freedoms, being able to provide for one's livelihood, 
or prevent authoritarianism, and this may lead to localized waves of disorders, with a diverse 
composition of participants. Finally, our work underscores the need to align citizen trust with governmental decisions,
especially at the local level, so that the implementation of
restrictions and other public health measures are perceived to be temporary and in the interest
of the common good, and that appropriate supporting policies
are promoted to ensure the least economic disruption and uncertainty.

\section{Data availability}
All datasets used in this study are available from ACLED \cite{RaleighIntroducingACLEDArmed2010}.
The source codes used are publicly available at {\href{https://github.com/gcampede/covid19-protests}{https://github.com/gcampede/covid19-protests} }.

\section{Acknowledgments}

MRD acknowledges support from the 
Army Research Office (W911NF-18-1-0345), and
the National Science Foundation (DMS-1814090).

\color{black}
\newpage

\printbibliography

@misc{ACLEDCOVID19DisorderTracker2020,
	title = {{COVID}-19 {Disorder} {Tracker}},
	url = {https://acleddata.com/analysis/covid-19-disorder-tracker/},
	language = {en-US},
	urldate = {2021-02-20},
	author = {ACLED},
	month = mar,
	year = {2020}
}

@article{MillschallengesdistributingCOVID192021,
	title = {The challenges of distributing {COVID}-19 vaccinations},
	volume = {31},
	issn = {2589-5370},
	url = {https://www.thelancet.com/journals/eclinm/article/PIIS2589-5370(20)30418-1/abstract},
	doi = {10.1016/j.eclinm.2020.100674},
	language = {English},
	urldate = {2021-02-22},
	journal = {EClinicalMedicine},
	author = {Mills, Melinda C. and Salisbury, David},
	month = jan,
	year = {2021},
	note = {Publisher: Elsevier}
}

@article{MurielVaccineDistributionEquity2021,
	title = {Vaccine {Distribution}—{Equity} {Left} {Behind}?},
	issn = {0098-7484},
	url = {https://doi.org/10.1001/jama.2021.1205},
	doi = {10.1001/jama.2021.1205},
	urldate = {2021-02-22},
	journal = {JAMA},
	author = {Muriel, Jean-Jacques and Bauchner, Howard},
	month = jan,
	year = {2021}
}

@article{ChuangLocalalliancesrivalries2019a,
	title = {Local alliances and rivalries shape near-repeat terror activity of al-{Qaeda}, {ISIS}, and insurgents},
	volume = {116},
	copyright = {Copyright © 2019 the Author(s). Published by PNAS.. https://creativecommons.org/licenses/by-nc-nd/4.0/This open access article is distributed under Creative Commons Attribution-NonCommercial-NoDerivatives License 4.0 (CC BY-NC-ND).},
	issn = {0027-8424, 1091-6490},
	url = {https://www.pnas.org/content/116/42/20898},
	doi = {10.1073/pnas.1904418116},
	language = {en},
	number = {42},
	urldate = {2021-04-23},
	journal = {Proceedings of the National Academy of Sciences},
	author = {Chuang, Yao-Li and Ben-Asher, Noam and D’Orsogna, Maria R.},
	month = oct,
	year = {2019},
	pmid = {31570597},
	note = {Publisher: National Academy of Sciences
Section: Social Sciences},
	pages = {20898--20903}
}

@article{OgataStatisticalModelsEarthquake1988b,
	title = {Statistical {Models} for {Earthquake} {Occurrences} and {Residual} {Analysis} for {Point} {Processes}},
	volume = {83},
	issn = {0162-1459},
	url = {https://www.jstor.org/stable/2288914},
	doi = {10.2307/2288914},
	number = {401},
	urldate = {2021-04-23},
	journal = {Journal of the American Statistical Association},
	author = {Ogata, Yosihiko},
	year = {1988},
	note = {Publisher: [American Statistical Association, Taylor \& Francis, Ltd.]},
	pages = {9--27}
}

@article{PapangelouIntegrabilityExpectedIncrements1972a,
	title = {Integrability of {Expected} {Increments} of {Point} {Processes} and a {Related} {Random} {Change} of {Scale}},
	volume = {165},
	issn = {0002-9947},
	url = {https://www.jstor.org/stable/1995899},
	doi = {10.2307/1995899},
	urldate = {2021-04-23},
	journal = {Transactions of the American Mathematical Society},
	author = {Papangelou, F.},
	year = {1972},
	note = {Publisher: American Mathematical Society},
	pages = {483--506}
}

@book{OConnorPracticalreliabilityengineering2012,
	address = {Chicester},
	edition = {5th ed},
	title = {Practical reliability engineering},
	isbn = {978-0-470-97981-5 978-0-470-97982-2},
	language = {eng},
	publisher = {Wiley},
	author = {O'Connor, Patrick D. T. and Kleyner, Andre},
	year = {2012},
	note = {OCLC: 775100516}
}

@book{ZarBiostatisticalAnalysis2010,
	title = {Biostatistical {Analysis}},
	isbn = {978-0-13-100846-5},
	language = {en},
	publisher = {Prentice Hall},
	author = {Zar, Jerrold H.},
	year = {2010}
}

@article{HilsenrathEthicsEconomicsCOVID192020,
	title = {Ethics and {Economics} of the {COVID}-19 {Pandemic} in the {United} {States}},
	volume = {7},
	issn = {2333-3928},
	url = {https://doi.org/10.1177/2333392820957661},
	doi = {10.1177/2333392820957661},
	language = {en},
	urldate = {2021-04-23},
	journal = {Health Services Research and Managerial Epidemiology},
	author = {Hilsenrath, Peter and Borders, Tyrone},
	month = jan,
	year = {2020},
	note = {Publisher: SAGE Publications Inc},
	pages = {2333392820957661}
}

@article{Al-DmourInfluenceSocialMedia2020,
	title = {Influence of {Social} {Media} {Platforms} on {Public} {Health} {Protection} {Against} the {COVID}-19 {Pandemic} via the {Mediating} {Effects} of {Public} {Health} {Awareness} and {Behavioral} {Changes}: {Integrated} {Model}},
	volume = {22},
	shorttitle = {Influence of {Social} {Media} {Platforms} on {Public} {Health} {Protection} {Against} the {COVID}-19 {Pandemic} via the {Mediating} {Effects} of {Public} {Health} {Awareness} and {Behavioral} {Changes}},
	url = {https://www.jmir.org/2020/8/e19996/},
	doi = {10.2196/19996},
	language = {en},
	number = {8},
	urldate = {2021-02-22},
	journal = {Journal of Medical Internet Research},
	author = {Al-Dmour, Hani and Masa’deh, Ra’ed and Salman, Amer and Abuhashesh, Mohammad and Al-Dmour, Rand},
	year = {2020},
	note = {Company: Journal of Medical Internet Research
Distributor: Journal of Medical Internet Research
Institution: Journal of Medical Internet Research
Label: Journal of Medical Internet Research
Publisher: JMIR Publications Inc., Toronto, Canada},
	pages = {e19996}
}

@article{ChenUnpackingblackbox2020,
	title = {Unpacking the black box: {How} to promote citizen engagement through government social media during the {COVID}-19 crisis},
	volume = {110},
	issn = {0747-5632},
	shorttitle = {Unpacking the black box},
	url = {https://www.sciencedirect.com/science/article/pii/S0747563220301333},
	doi = {10.1016/j.chb.2020.106380},
	language = {en},
	urldate = {2021-02-22},
	journal = {Computers in Human Behavior},
	author = {Chen, Qiang and Min, Chen and Zhang, Wei and Wang, Ge and Ma, Xiaoyue and Evans, Richard},
	month = sep,
	year = {2020},
	pages = {106380}
}

@misc{DatareportalDigital2021Mexico2021,
	title = {Digital 2021: {Mexico}},
	url = {https://datareportal.com/reports/digital-2021-mexico},
	author = {Datareportal},
	year = {2021}
}

@misc{DatareportalDigital2020India2020,
	title = {Digital 2020: {India}},
	url = {https://datareportal.com/reports/digital-2020-india},
	author = {Datareportal},
	year = {2020}
}

@misc{DatareportalDigital2020Israel2020,
	title = {Digital 2020: {Israel}},
	url = {https://datareportal.com/reports/digital-2020-israel},
	author = {Datareportal},
	year = {2020}
}

@article{Jaffe-HoffmanRedlightgreen2020,
	title = {Red light, green light, go! {Gamzu}’s traffic light plan passes},
	url = {https://www.jpost.com/health-science/coronavirus-cabinet-to-convene-sneak-peek-at-gamzus-traffic-light-plan-640464},
	journal = {The Jerusalem Post},
	author = {Jaffe-Hoffman, Maayan},
	year = {2020}
}

@article{ShpigelClashesarrestshundreds2020,
	title = {Clashes, arrests as hundreds of anti-{Netanyahu} protests held across {Israel} under lockdown},
	url = {https://www.haaretz.com/israel-news/.premium-despite-restrictions-anti-netanyahu-protests-continue-at-thousands-of-locations-1.9205689},
	language = {en},
	urldate = {2021-02-22},
	journal = {Haaretz},
	author = {Shpigel, Noa and Peleg, Bar and Hasson, Nir and Breiner, Josh and Shezaf, Hagar},
	year = {2020}
}

@article{MurrayMurderswomenMexico2020,
	title = {Murders of women in {Mexico} rise amid fears of lockdown violence},
	url = {https://www.reuters.com/article/us-mexico-women-violence-trfn-idUSKCN22930V},
	language = {en},
	urldate = {2021-02-22},
	journal = {Reuters},
	author = {Murray, Christine, Oscar Lopez},
	month = apr,
	year = {2020}
}

@article{SanchezMexicofocusescoronavirus2020,
	title = {As {Mexico} focuses on coronavirus, drug gang violence rises},
	url = {https://www.reuters.com/article/uk-health-coronavirus-mexico-cartels-idUKKBN23P1T5},
	language = {en},
	urldate = {2021-02-24},
	journal = {Reuters},
	author = {Sanchez, Uriel, Drazen Jorgic},
	month = jun,
	year = {2020}
}

@article{PennycookFightingCOVID19Misinformation2020,
	title = {Fighting {COVID}-19 {Misinformation} on {Social} {Media}: {Experimental} {Evidence} for a {Scalable} {Accuracy}-{Nudge} {Intervention}},
	volume = {31},
	issn = {0956-7976, 1467-9280},
	shorttitle = {Fighting {COVID}-19 {Misinformation} on {Social} {Media}},
	url = {http://journals.sagepub.com/doi/10.1177/0956797620939054},
	doi = {10.1177/0956797620939054},
	language = {en},
	number = {7},
	urldate = {2021-02-24},
	journal = {Psychological Science},
	author = {Pennycook, Gordon and McPhetres, Jonathon and Zhang, Yunhao and Lu, Jackson G. and Rand, David G.},
	month = jul,
	year = {2020},
	pages = {770--780}
}

@article{TasnimImpactRumorsMisinformation2020,
	title = {Impact of {Rumors} and {Misinformation} on {COVID}-19 in {Social} {Media}},
	volume = {53},
	issn = {1975-8375, 2233-4521},
	url = {http://jpmph.org/journal/view.php?doi=10.3961/jpmph.20.094},
	doi = {10.3961/jpmph.20.094},
	language = {en},
	number = {3},
	urldate = {2021-02-24},
	journal = {Journal of Preventive Medicine and Public Health},
	author = {Tasnim, Samia and Hossain, Md Mahbub and Mazumder, Hoimonty},
	month = may,
	year = {2020},
	pages = {171--174}
}

@article{KouzyCoronavirusGoesViral2020,
	title = {Coronavirus {Goes} {Viral}: {Quantifying} the {COVID}-19 {Misinformation} {Epidemic} on {Twitter}},
	volume = {12},
	issn = {2168-8184},
	shorttitle = {Coronavirus {Goes} {Viral}},
	url = {https://www.ncbi.nlm.nih.gov/pmc/articles/PMC7152572/},
	doi = {10.7759/cureus.7255},
	number = {3},
	urldate = {2021-02-24},
	journal = {Cureus},
	author = {Kouzy, Ramez and Abi Jaoude, Joseph and Kraitem, Afif and El Alam, Molly B and Karam, Basil and Adib, Elio and Zarka, Jabra and Traboulsi, Cindy and Akl, Elie W and Baddour, Khalil},
	year = {2020},
	pmid = {null},
	pmcid = {PMC7152572}
}

@techreport{BradyFragmentedUnitedStates2020,
	address = {Rochester, NY},
	type = {{SSRN} {Scholarly} {Paper}},
	title = {The {Fragmented} {United} {States} of {America}: {The} {Impact} of {Scattered} {Lock}-{Down} {Policies} on {Country}-{Wide} {Infections}},
	shorttitle = {The {Fragmented} {United} {States} of {America}},
	url = {https://papers.ssrn.com/abstract=3681486},
	language = {en},
	number = {ID 3681486},
	urldate = {2021-02-24},
	institution = {Social Science Research Network},
	author = {Brady, Ryan Robert and Insler, Mike and Rothert, Jacek},
	month = aug,
	year = {2020},
	doi = {10.2139/ssrn.3681486}
}

@article{AlfanoEfficacyLockdownCOVID192020,
	title = {The {Efficacy} of {Lockdown} {Against} {COVID}-19: {A} {Cross}-{Country} {Panel} {Analysis}},
	volume = {18},
	issn = {1179-1896},
	shorttitle = {The {Efficacy} of {Lockdown} {Against} {COVID}-19},
	doi = {10.1007/s40258-020-00596-3},
	language = {eng},
	number = {4},
	journal = {Applied Health Economics and Health Policy},
	author = {Alfano, Vincenzo and Ercolano, Salvatore},
	month = aug,
	year = {2020},
	pmid = {32495067},
	pmcid = {PMC7268966},
	pages = {509--517}
}

@article{DelenNoPlaceHome2020,
	title = {No {Place} {Like} {Home}: {Cross}-{National} {Data} {Analysis} of the {Efficacy} of {Social} {Distancing} {During} the {COVID}-19 {Pandemic}},
	volume = {6},
	issn = {2369-2960},
	shorttitle = {No {Place} {Like} {Home}},
	url = {http://publichealth.jmir.org/2020/2/e19862/},
	doi = {10.2196/19862},
	language = {en},
	number = {2},
	urldate = {2021-02-24},
	journal = {JMIR Public Health and Surveillance},
	author = {Delen, Dursun and Eryarsoy, Enes and Davazdahemami, Behrooz},
	month = may,
	year = {2020},
	pages = {e19862}
}

@article{WelleniusImpactsUSStateLevel2020,
	title = {Impacts of {US} {State}-{Level} {Social} {Distancing} {Policies} on {Population} {Mobility} and {COVID}-19 {Case} {Growth} {During} the {First} {Wave} of the {Pandemic}},
	url = {http://arxiv.org/abs/2004.10172},
	urldate = {2021-02-24},
	journal = {arXiv:2004.10172 [q-bio]},
	author = {Wellenius, Gregory A. and Vispute, Swapnil and Espinosa, Valeria and Fabrikant, Alex and Tsai, Thomas C. and Hennessy, Jonathan and Dai, Andrew and Williams, Brian and Gadepalli, Krishna and Boulanger, Adam and Pearce, Adam and Kamath, Chaitanya and Schlosberg, Arran and Bendebury, Catherine and Mandayam, Chinmoy and Stanton, Charlotte and Bavadekar, Shailesh and Pluntke, Christopher and Desfontaines, Damien and Jacobson, Benjamin and Armstrong, Zan and Gipson, Bryant and Wilson, Royce and Widdowson, Andrew and Chou, Katherine and Oplinger, Andrew and Shekel, Tomer and Jha, Ashish K. and Gabrilovich, Evgeniy},
	month = nov,
	year = {2020},
	note = {arXiv: 2004.10172}
}

@article{GaleaCOVID19PandemicUnemployment2020,
	title = {{COVID}-19 {Pandemic}, {Unemployment}, and {Civil} {Unrest}: {Underlying} {Deep} {Racial} and {Socioeconomic} {Divides}},
	volume = {324},
	issn = {0098-7484},
	shorttitle = {{COVID}-19 {Pandemic}, {Unemployment}, and {Civil} {Unrest}},
	url = {https://jamanetwork.com/journals/jama/fullarticle/2767354},
	doi = {10.1001/jama.2020.11132},
	language = {en},
	number = {3},
	urldate = {2020-10-22},
	journal = {JAMA},
	author = {Galea, Sandro and Abdalla, Salma M.},
	month = jul,
	year = {2020},
	pages = {227}
}

@article{Bonnasse-GahotEpidemiologicalmodelling20052018,
	title = {Epidemiological modelling of the 2005 {French} riots: a spreading wave and the role of contagion},
	volume = {8},
	copyright = {2018 The Author(s)},
	issn = {2045-2322},
	shorttitle = {Epidemiological modelling of the 2005 {French} riots},
	url = {https://www.nature.com/articles/s41598-017-18093-4},
	doi = {10.1038/s41598-017-18093-4},
	language = {en},
	number = {1},
	urldate = {2020-10-22},
	journal = {Scientific Reports},
	author = {Bonnasse-Gahot, Laurent and Berestycki, Henri and Depuiset, Marie-Aude and Gordon, Mirta B. and Roché, Sebastian and Rodriguez, Nancy and Nadal, Jean-Pierre},
	month = jan,
	year = {2018},
	note = {Number: 1
Publisher: Nature Publishing Group},
	pages = {107}
}

@article{Sullivancriticalmasscrowd1977,
	title = {The critical mass in crowd behavior: {Crowd} size, contagion and the evolution of riots},
	volume = {4},
	issn = {0160-4341(Print)},
	shorttitle = {The critical mass in crowd behavior},
	number = {2},
	journal = {Humboldt Journal of Social Relations},
	author = {Sullivan, Thomas J.},
	year = {1977},
	note = {Place: US
Publisher: Humboldt State University},
	pages = {46--59}
}

@article{BaudainsGeographicpatternsdiffusion2013,
	title = {Geographic patterns of diffusion in the 2011 {London} riots},
	volume = {45},
	issn = {0143-6228},
	url = {http://www.sciencedirect.com/science/article/pii/S0143622813002154},
	doi = {10.1016/j.apgeog.2013.09.010},
	language = {en},
	urldate = {2020-10-22},
	journal = {Applied Geography},
	author = {Baudains, Peter and Johnson, Shane D. and Braithwaite, Alex Maves},
	month = dec,
	year = {2013},
	pages = {211--219}
}

@article{BaudainsTargetChoiceExtreme2013,
	title = {Target {Choice} {During} {Extreme} {Events}: {A} {Discrete} {Spatial} {Choice} {Model} of the 2011 {London} {Riots}},
	volume = {51},
	copyright = {© 2013 American Society of Criminology},
	issn = {1745-9125},
	shorttitle = {Target {Choice} {During} {Extreme} {Events}},
	url = {https://onlinelibrary.wiley.com/doi/abs/10.1111/1745-9125.12004},
	doi = {10.1111/1745-9125.12004},
	language = {en},
	number = {2},
	urldate = {2020-10-22},
	journal = {Criminology},
	author = {Baudains, Peter and Braithwaite, Alex and Johnson, Shane D.},
	year = {2013},
	note = {\_eprint: https://onlinelibrary.wiley.com/doi/pdf/10.1111/1745-9125.12004},
	pages = {251--285}
}

@article{MidlarskyAnalyzingDiffusionContagion1978a,
	title = {Analyzing {Diffusion} and {Contagion} {Effects}: {The} {Urban} {Disorders} of the 1960s},
	volume = {72},
	issn = {0003-0554},
	shorttitle = {Analyzing {Diffusion} and {Contagion} {Effects}},
	url = {https://www.jstor.org/stable/1955117},
	doi = {10.2307/1955117},
	number = {3},
	urldate = {2020-10-22},
	journal = {The American Political Science Review},
	author = {Midlarsky, Manus I.},
	year = {1978},
	note = {Publisher: [American Political Science Association, Cambridge University Press]},
	pages = {996--1008}
}

@article{Berestyckimodelriotsdynamics2015,
	title = {A model of riots dynamics: shocks, diffusion and thresholds},
	shorttitle = {A model of riots dynamics},
	url = {http://arxiv.org/abs/1502.04725},
	urldate = {2020-10-22},
	journal = {arXiv:1502.04725 [physics]},
	author = {Berestycki, Henri and Nadal, Jean-Pierre and Rodriguez, Nancy},
	month = jul,
	year = {2015},
	note = {arXiv: 1502.04725}
}

@article{HawkesSpectraselfexcitingmutually1971a,
	title = {Spectra of some self-exciting and mutually exciting point processes},
	volume = {58},
	issn = {0006-3444},
	url = {https://academic.oup.com/biomet/article/58/1/83/224809},
	doi = {10.1093/biomet/58.1.83},
	language = {en},
	number = {1},
	urldate = {2020-10-22},
	journal = {Biometrika},
	author = {Hawkes, Alan G.},
	month = apr,
	year = {1971},
	note = {Publisher: Oxford Academic},
	pages = {83--90}
}

@techreport{PengMultidimensionalPointProcess2002,
	title = {Multi-dimensional {Point} {Process} {Models} in {R}},
	url = {https://escholarship.org/content/qt3n6609wb/qt3n6609wb.pdf?t=lnp7c3},
	institution = {UCLA},
	author = {Peng, Roger D.},
	year = {2002}
}

@book{HastieElementsStatisticalLearning2013a,
	address = {New York, NY},
	edition = {2 edizione},
	title = {The {Elements} of {Statistical} {Learning}: {Data} {Mining}, {Inference}, and {Prediction}},
	isbn = {978-0-387-84857-0},
	shorttitle = {The {Elements} of {Statistical} {Learning}},
	language = {Inglese},
	publisher = {Springer Nature},
	author = {Hastie, Trevor and Tibshirani, Robert and Friedman, Jerome},
	month = jun,
	year = {2013}
}

@techreport{ACLEDArmedConflictLocation2019,
	title = {Armed {Conflict} {Location}	\& {Event} {Data} {Project} ({ACLED}) {Codebook}},
	url = {https://acleddata.com/acleddatanew/wp-content/uploads/dlm_uploads/2019/04/ACLED_Codebook_2019FINAL_pbl.pdf},
	author = {ACLED},
	year = {2019}
}

@article{KingHighTimesHate2013,
	title = {High {Times} for {Hate} {Crimes}: {Explaining} the {Temporal} {Clustering} of {Hate}-{Motivated} {Offending}},
	volume = {51},
	copyright = {© 2013 American Society of Criminology},
	issn = {1745-9125},
	shorttitle = {High {Times} for {Hate} {Crimes}},
	url = {https://onlinelibrary.wiley.com/doi/abs/10.1111/1745-9125.12022},
	doi = {https://doi.org/10.1111/1745-9125.12022},
	language = {en},
	number = {4},
	urldate = {2020-11-29},
	journal = {Criminology},
	author = {King, Ryan D. and Sutton, Gretchen M.},
	year = {2013},
	note = {\_eprint: https://onlinelibrary.wiley.com/doi/pdf/10.1111/1745-9125.12022},
	pages = {871--894}
}

@book{DellaPortaSocialMovementsIntroduction2020,
	title = {Social {Movements}: {An} {Introduction}},
	isbn = {978-1-119-16765-5},
	shorttitle = {Social {Movements}},
	language = {en},
	publisher = {John Wiley \& Sons},
	author = {Della Porta, Donatella and Diani, Mario},
	month = apr,
	year = {2020}
}

@article{Drurysocialidentitymodel2020,
	title = {A social identity model of riot diffusion: {From} injustice to empowerment in the 2011 {London} riots},
	volume = {50},
	copyright = {© 2019 John Wiley \& Sons, Ltd.},
	issn = {1099-0992},
	shorttitle = {A social identity model of riot diffusion},
	url = {https://onlinelibrary.wiley.com/doi/abs/10.1002/ejsp.2650},
	doi = {https://doi.org/10.1002/ejsp.2650},
	language = {en},
	number = {3},
	urldate = {2020-11-29},
	journal = {European Journal of Social Psychology},
	author = {Drury, John and Stott, Clifford and Ball, Roger and Reicher, Stephen and Neville, Fergus and Bell, Linda and Biddlestone, Mikey and Choudhury, Sanjeedah and Lovell, Max and Ryan, Caoimhe},
	year = {2020},
	note = {\_eprint: https://onlinelibrary.wiley.com/doi/pdf/10.1002/ejsp.2650},
	pages = {646--661}
}

@article{MidlarskyWhyViolenceSpreads1980c,
	title = {Why {Violence} {Spreads}: {The} {Contagion} of {International} {Terrorism}},
	volume = {24},
	issn = {0020-8833},
	shorttitle = {Why {Violence} {Spreads}},
	url = {https://www.jstor.org/stable/2600202},
	doi = {10.2307/2600202},
	number = {2},
	urldate = {2020-11-29},
	journal = {International Studies Quarterly},
	author = {Midlarsky, Manus I. and Crenshaw, Martha and Yoshida, Fumihiko},
	year = {1980},
	note = {Publisher: [International Studies Association, Wiley]},
	pages = {262--298}
}

@article{DaviesmathematicalmodelLondon2013,
	title = {A mathematical model of the {London} riots and their policing},
	volume = {3},
	copyright = {2013 The Author(s)},
	issn = {2045-2322},
	url = {https://www.nature.com/articles/srep01303},
	doi = {10.1038/srep01303},
	language = {en},
	number = {1},
	urldate = {2020-11-30},
	journal = {Scientific Reports},
	author = {Davies, Toby P. and Fry, Hannah M. and Wilson, Alan G. and Bishop, Steven R.},
	month = feb,
	year = {2013},
	note = {Number: 1
Publisher: Nature Publishing Group},
	pages = {1303}
}

@article{CadenaForecastingSocialUnrest2015,
	title = {Forecasting {Social} {Unrest} {Using} {Activity} {Cascades}},
	volume = {10},
	issn = {1932-6203},
	url = {https://journals.plos.org/plosone/article?id=10.1371/journal.pone.0128879},
	doi = {10.1371/journal.pone.0128879},
	language = {en},
	number = {6},
	urldate = {2020-11-30},
	journal = {PLOS ONE},
	author = {Cadena, Jose and Korkmaz, Gizem and Kuhlman, Chris J. and Marathe, Achla and Ramakrishnan, Naren and Vullikanti, Anil},
	month = jun,
	year = {2015},
	note = {Publisher: Public Library of Science},
	pages = {e0128879}
}

@article{HussainWhatBestExplains2013,
	title = {What {Best} {Explains} {Successful} {Protest} {Cascades}? {ICTs} and the {Fuzzy} {Causes} of the {Arab} {Spring}},
	volume = {15},
	issn = {1521-9488},
	shorttitle = {What {Best} {Explains} {Successful} {Protest} {Cascades}?},
	url = {https://academic.oup.com/isr/article/15/1/48/1792866},
	doi = {10.1111/misr.12020},
	language = {en},
	number = {1},
	urldate = {2020-11-30},
	journal = {International Studies Review},
	author = {Hussain, Muzammil M. and Howard, Philip N.},
	month = mar,
	year = {2013},
	note = {Publisher: Oxford Academic},
	pages = {48--66}
}

@article{HaleRegimeChangeCascades2013,
	title = {Regime {Change} {Cascades}: {What} {We} {Have} {Learned} from the 1848 {Revolutions} to the 2011 {Arab} {Uprisings}},
	volume = {16},
	shorttitle = {Regime {Change} {Cascades}},
	url = {https://doi.org/10.1146/annurev-polisci-032211-212204},
	doi = {10.1146/annurev-polisci-032211-212204},
	number = {1},
	urldate = {2020-11-30},
	journal = {Annual Review of Political Science},
	author = {Hale, Henry E.},
	year = {2013},
	note = {\_eprint: https://doi.org/10.1146/annurev-polisci-032211-212204},
	pages = {331--353}
}

@book{McCallIdentitiesInteractionsExamination1978,
	title = {Identities and {Interactions}: {An} {Examination} of {Human} {Associations} in {Everyday} {Life}},
	isbn = {978-0-02-920630-0},
	shorttitle = {Identities and {Interactions}},
	language = {en},
	publisher = {Free Press},
	author = {McCall, George J. and Simmons, Jerry Laird},
	year = {1978}
}

@article{RaleighIntroducingACLEDArmed2010,
	title = {Introducing {ACLED}: {An} {Armed} {Conflict} {Location} and {Event} {Dataset}: {Special} {Data} {Feature}},
	copyright = {© The Author(s) 2010},
	shorttitle = {Introducing {ACLED}},
	url = {https://journals.sagepub.com/doi/10.1177/0022343310378914},
	doi = {10.1177/0022343310378914},
	language = {en},
	urldate = {2020-11-30},
	journal = {Journal of Peace Research},
	author = {Raleigh, Clionadh and Linke, Andrew and Hegre, Håvard and Karlsen, Joakim},
	month = sep,
	year = {2010},
	note = {Publisher: SAGE PublicationsSage UK: London, England}
}

@article{LloydLeastsquaresquantization1982a,
	title = {Least squares quantization in {PCM}},
	volume = {28},
	issn = {1557-9654},
	doi = {10.1109/TIT.1982.1056489},
	number = {2},
	journal = {IEEE Transactions on Information Theory},
	author = {Lloyd, S.},
	month = mar,
	year = {1982},
	note = {Conference Name: IEEE Transactions on Information Theory},
	pages = {129--137}
}

@article{TenchSpatiotemporalpatternsIED2016g,
	title = {Spatio-temporal patterns of {IED} usage by the {Provisional} {Irish} {Republican} {Army}},
	volume = {27},
	copyright = {open},
	issn = {1469-4425},
	url = {http://dx.doi.org/10.1017/S0956792515000686},
	language = {eng},
	number = {3},
	urldate = {2020-11-30},
	journal = {European Journal of Applied Mathematics},
	author = {Tench, S. and Fry, H. and Gill, P.},
	month = jun,
	year = {2016},
	note = {Number: 3},
	pages = {377--402}
}

@article{GaoMentalhealthproblems2020a,
	title = {Mental health problems and social media exposure during {COVID}-19 outbreak},
	volume = {15},
	issn = {1932-6203},
	url = {https://dx.plos.org/10.1371/journal.pone.0231924},
	doi = {10.1371/journal.pone.0231924},
	language = {en},
	number = {4},
	urldate = {2020-11-30},
	journal = {PLOS ONE},
	author = {Gao, Junling and Zheng, Pinpin and Jia, Yingnan and Chen, Hao and Mao, Yimeng and Chen, Suhong and Wang, Yi and Fu, Hua and Dai, Junming},
	editor = {Hashimoto, Kenji},
	month = apr,
	year = {2020},
	pages = {e0231924}
}

@article{DornCOVID19exacerbatinginequalities2020,
	title = {{COVID}-19 exacerbating inequalities in the {US}},
	volume = {395},
	issn = {0140-6736},
	url = {https://www.ncbi.nlm.nih.gov/pmc/articles/PMC7162639/},
	doi = {10.1016/S0140-6736(20)30893-X},
	number = {10232},
	urldate = {2020-11-30},
	journal = {Lancet (London, England)},
	author = {Dorn, Aaron van and Cooney, Rebecca E and Sabin, Miriam L},
	year = {2020},
	pmid = {32305087},
	pmcid = {PMC7162639},
	pages = {1243--1244}
}

@article{PatelPovertyinequalityCOVID192020,
	title = {Poverty, inequality and {COVID}-19: the forgotten vulnerable},
	volume = {183},
	issn = {0033-3506},
	shorttitle = {Poverty, inequality and {COVID}-19},
	url = {https://www.ncbi.nlm.nih.gov/pmc/articles/PMC7221360/},
	doi = {10.1016/j.puhe.2020.05.006},
	urldate = {2020-11-30},
	journal = {Public Health},
	author = {Patel, J.A. and Nielsen, F.B.H. and Badiani, A.A. and Assi, S. and Unadkat, V.A. and Patel, B. and Ravindrane, R. and Wardle, H.},
	month = jun,
	year = {2020},
	pmid = {32502699},
	pmcid = {PMC7221360},
	pages = {110--111}
}

@article{SibleyEffectsCOVID19pandemic2020,
	title = {Effects of the {COVID}-19 pandemic and nationwide lockdown on trust, attitudes toward government, and well-being.},
	volume = {75},
	issn = {1935-990X},
	url = {https://psycnet.apa.org/fulltext/2020-39514-001.pdf},
	doi = {10.1037/amp0000662},
	number = {5},
	urldate = {2020-11-30},
	journal = {American Psychologist},
	author = {Sibley, Chris G. and Greaves, Lara M. and Satherley, Nicole and Wilson, Marc S. and Overall, Nickola C. and Lee, Carol H. J. and Milojev, Petar and Bulbulia, Joseph and Osborne, Danny and Milfont, Taciano L. and Houkamau, Carla A. and Duck, Isabelle M. and Vickers-Jones, Raine and Barlow, Fiona Kate},
	year = {2020},
	note = {Publisher: US: American Psychological Association},
	pages = {618}
}

@book{ClementPeopleHistoryRiots2016,
	address = {London},
	title = {A {People}’s {History} of {Riots}, {Protest} and the {Law}},
	isbn = {978-1-137-52750-9 978-1-137-52751-6},
	url = {http://link.springer.com/10.1057/978-1-137-52751-6},
	language = {en},
	urldate = {2020-11-30},
	publisher = {Palgrave Macmillan UK},
	author = {Clement, Matt},
	year = {2016},
	doi = {10.1057/978-1-137-52751-6}
}

@book{TrottierSocialMediaPolitics2014,
	title = {Social {Media}, {Politics} and the {State}: {Protests}, {Revolutions}, {Riots}, {Crime} and {Policing} in the {Age} of {Facebook}, {Twitter} and {YouTube}},
	isbn = {978-1-317-65548-0},
	shorttitle = {Social {Media}, {Politics} and the {State}},
	language = {en},
	publisher = {Routledge},
	author = {Trottier, Daniel and Fuchs, Christian},
	month = jul,
	year = {2014}
}

@article{JostHowSocialMedia2018,
	title = {How {Social} {Media} {Facilitates} {Political} {Protest}: {Information}, {Motivation}, and {Social} {Networks}},
	volume = {39},
	copyright = {© 2018 International Society of Political Psychology},
	issn = {1467-9221},
	shorttitle = {How {Social} {Media} {Facilitates} {Political} {Protest}},
	url = {https://onlinelibrary.wiley.com/doi/abs/10.1111/pops.12478},
	doi = {https://doi.org/10.1111/pops.12478},
	language = {en},
	number = {S1},
	urldate = {2020-11-30},
	journal = {Political Psychology},
	author = {Jost, John T. and Barberá, Pablo and Bonneau, Richard and Langer, Melanie and Metzger, Megan and Nagler, Jonathan and Sterling, Joanna and Tucker, Joshua A.},
	year = {2018},
	note = {\_eprint: https://onlinelibrary.wiley.com/doi/pdf/10.1111/pops.12478},
	pages = {85--118}
}

@article{PoellSocialmediatransformation2014,
	title = {Social media and the transformation of activist communication: exploring the social media ecology of the 2010 {Toronto} {G20} protests},
	volume = {17},
	issn = {1369-118X},
	shorttitle = {Social media and the transformation of activist communication},
	url = {https://doi.org/10.1080/1369118X.2013.812674},
	doi = {10.1080/1369118X.2013.812674},
	number = {6},
	urldate = {2020-11-30},
	journal = {Information, Communication \& Society},
	author = {Poell, Thomas},
	month = jul,
	year = {2014},
	note = {Publisher: Routledge
\_eprint: https://doi.org/10.1080/1369118X.2013.812674},
	pages = {716--731}
}

@article{GreerWePredictRiot2010,
	title = {We {Predict} a {Riot}?{Public} {Order} {Policing}, {New} {Media} {Environments} and the {Rise} of the {Citizen} {Journalist}},
	volume = {50},
	issn = {0007-0955},
	shorttitle = {We {Predict} a {Riot}?},
	url = {https://academic.oup.com/bjc/article/50/6/1041/405457},
	doi = {10.1093/bjc/azq039},
	language = {en},
	number = {6},
	urldate = {2020-11-30},
	journal = {The British Journal of Criminology},
	author = {Greer, Chris and McLaughlin, Eugene},
	month = nov,
	year = {2010},
	note = {Publisher: Oxford Academic},
	pages = {1041--1059}
}

@article{ThomasWhenWillCollective2014,
	title = {When {Will} {Collective} {Action} {Be} {Effective}? {Violent} and {Non}-{Violent} {Protests} {Differentially} {Influence} {Perceptions} of {Legitimacy} and {Efficacy} {Among} {Sympathizers} ,  {When} {Will} {Collective} {Action} {Be} {Effective}? {Violent} and {Non}-{Violent} {Protests} {Differentially} {Influence} {Perceptions} of {Legitimacy} and {Efficacy} {Among} {Sympathizers}},
	volume = {40},
	issn = {0146-1672},
	shorttitle = {When {Will} {Collective} {Action} {Be} {Effective}?},
	url = {https://doi.org/10.1177/0146167213510525},
	doi = {10.1177/0146167213510525},
	language = {en},
	number = {2},
	urldate = {2020-11-30},
	journal = {Personality and Social Psychology Bulletin},
	author = {Thomas, Emma F. and Louis, Winnifred R.},
	month = feb,
	year = {2014},
	note = {Publisher: SAGE Publications Inc},
	pages = {263--276}
}

@article{GeschwenderCivilRightsProtest1968,
	title = {Civil {Rights} {Protest} and {Riots}: {A} {Disappearing} {Distinction}},
	volume = {49},
	issn = {0038-4941},
	shorttitle = {Civil {Rights} {Protest} and {Riots}},
	url = {https://www.jstor.org/stable/42858409},
	number = {3},
	urldate = {2020-11-30},
	journal = {Social Science Quarterly},
	author = {Geschwender, James A.},
	year = {1968},
	note = {Publisher: [University of Texas Press, Wiley]},
	pages = {474--484}
}

@article{MidlarskyMathematicalModelsInstability1970,
	title = {Mathematical {Models} of {Instability} and a {Theory} of {Diffusion}},
	volume = {14},
	issn = {00208833},
	url = {https://academic.oup.com/isq/article-lookup/doi/10.2307/3013540},
	doi = {10.2307/3013540},
	number = {1},
	urldate = {2020-11-30},
	journal = {International Studies Quarterly},
	author = {Midlarsky, Manus},
	month = mar,
	year = {1970},
	pages = {60}
}

@article{LiCoupContagionHypothesis1975,
	title = {The "{Coup} {Contagion}" {Hypothesis}},
	volume = {19},
	issn = {0022-0027},
	url = {https://doi.org/10.1177/002200277501900104},
	doi = {10.1177/002200277501900104},
	language = {en},
	number = {1},
	urldate = {2020-11-30},
	journal = {Journal of Conflict Resolution},
	author = {Li, Richard P.Y. and Thompson, William R.},
	month = mar,
	year = {1975},
	note = {Publisher: SAGE Publications Inc},
	pages = {63--84}
}

@article{BuhaugContagionConfusionWhy2008a,
	title = {Contagion or {Confusion}? {Why} {Conflicts} {Cluster} in {Space}},
	volume = {52},
	issn = {0020-8833},
	shorttitle = {Contagion or {Confusion}?},
	url = {https://academic.oup.com/isq/article/52/2/215/1792634},
	doi = {10.1111/j.1468-2478.2008.00499.x},
	language = {en},
	number = {2},
	urldate = {2020-11-30},
	journal = {International Studies Quarterly},
	author = {Buhaug, Halvard and Gleditsch, Kristian Skrede},
	month = jun,
	year = {2008},
	note = {Publisher: Oxford Academic},
	pages = {215--233}
}

@article{SchutteDiffusionpatternsviolence2011,
	title = {Diffusion patterns of violence in civil wars},
	volume = {30},
	issn = {0962-6298},
	url = {http://www.sciencedirect.com/science/article/pii/S0962629811000424},
	doi = {10.1016/j.polgeo.2011.03.005},
	language = {en},
	number = {3},
	urldate = {2020-11-30},
	journal = {Political Geography},
	author = {Schutte, Sebastian and Weidmann, Nils B.},
	month = mar,
	year = {2011},
	pages = {143--152}
}

@article{LewisSelfexcitingpointprocess2012e,
	title = {Self-exciting point process models of civilian deaths in {Iraq}},
	volume = {25},
	issn = {1743-4645},
	url = {https://doi.org/10.1057/sj.2011.21},
	doi = {10.1057/sj.2011.21},
	language = {en},
	number = {3},
	urldate = {2020-11-30},
	journal = {Security Journal},
	author = {Lewis, Erik and Mohler, George and Brantingham, P Jeffrey and Bertozzi, Andrea L},
	month = jul,
	year = {2012},
	pages = {244--264}
}

@article{MyersDiffusionCollectiveViolence2000a,
	title = {The {Diffusion} of {Collective} {Violence}: {Infectiousness}, {Susceptibility}, and {Mass} {Media} {Networks}},
	volume = {106},
	issn = {0002-9602},
	shorttitle = {The {Diffusion} of {Collective} {Violence}},
	url = {https://www.journals.uchicago.edu/doi/abs/10.1086/303110},
	doi = {10.1086/303110},
	number = {1},
	urldate = {2020-11-30},
	journal = {American Journal of Sociology},
	author = {Myers, Daniel J.},
	month = jul,
	year = {2000},
	note = {Publisher: The University of Chicago Press},
	pages = {173--208}
}

@article{HouleDiffusionConfusionClustered2016,
	title = {Diffusion or {Confusion}? {Clustered} {Shocks} and the {Conditional} {Diffusion} of {Democracy}},
	volume = {70},
	issn = {0020-8183, 1531-5088},
	shorttitle = {Diffusion or {Confusion}?},
	url = {https://www.cambridge.org/core/journals/international-organization/article/diffusion-or-confusion-clustered-shocks-and-the-conditional-diffusion-of-democracy/FAA262B7C0A8BA0FA7D69956FCB6A0EF},
	doi = {10.1017/S002081831600028X},
	language = {en},
	number = {4},
	urldate = {2020-11-30},
	journal = {International Organization},
	author = {Houle, Christian and Kayser, Mark A. and Xiang, Jun},
	year = {2016},
	note = {Publisher: Cambridge University Press},
	pages = {687--726}
}

@article{SimmonsGlobalizationLiberalizationPolicy2004,
	title = {The {Globalization} of {Liberalization}: {Policy} {Diffusion} in the {International} {Political} {Economy}},
	volume = {98},
	issn = {1537-5943, 0003-0554},
	shorttitle = {The {Globalization} of {Liberalization}},
	doi = {10.1017/S0003055404001078},
	language = {en},
	number = {1},
	urldate = {2020-11-30},
	journal = {American Political Science Review},
	author = {Simmons, Beth A. and Elkins, Zachary},
	month = feb,
	year = {2004},
	note = {Publisher: Cambridge University Press},
	pages = {171--189}
}

@article{ShortMeasuringModelingRepeat2009,
	title = {Measuring and {Modeling} {Repeat} and {Near}-{Repeat} {Burglary} {Effects}},
	volume = {25},
	issn = {1573-7799},
	url = {https://doi.org/10.1007/s10940-009-9068-8},
	doi = {10.1007/s10940-009-9068-8},
	language = {en},
	number = {3},
	urldate = {2020-11-30},
	journal = {Journal of Quantitative Criminology},
	author = {Short, M. B. and D’Orsogna, M. R. and Brantingham, P. J. and Tita, G. E.},
	month = sep,
	year = {2009},
	pages = {325--339}
}

@article{AndrewsFirstconfirmedcase2020,
	title = {First confirmed case of {COVID}-19 infection in {India}: {A} case report},
	volume = {151},
	issn = {0971-5916},
	shorttitle = {First confirmed case of {COVID}-19 infection in {India}},
	url = {https://www.ncbi.nlm.nih.gov/pmc/articles/PMC7530459/},
	doi = {10.4103/ijmr.IJMR_2131_20},
	number = {5},
	urldate = {2020-11-30},
	journal = {The Indian Journal of Medical Research},
	author = {Andrews, M.A. and Areekal, Binu and Rajesh, K.R. and Krishnan, Jijith and Suryakala, R. and Krishnan, Biju and Muraly, C.P. and Santhosh, P.V.},
	month = may,
	year = {2020},
	pmid = {32611918},
	pmcid = {PMC7530459},
	pages = {490--492}
}

@article{LastfirstwaveCOVID192020,
	title = {The first wave of {COVID}-19 in {Israel}—{Initial} analysis of publicly available data},
	volume = {15},
	issn = {1932-6203},
	url = {https://journals.plos.org/plosone/article?id=10.1371/journal.pone.0240393},
	doi = {10.1371/journal.pone.0240393},
	language = {en},
	number = {10},
	urldate = {2020-11-30},
	journal = {PLOS ONE},
	author = {Last, Mark},
	month = oct,
	year = {2020},
	note = {Publisher: Public Library of Science},
	pages = {e0240393}
}

@article{BaudainsSpatialPatterns20112013,
	title = {Spatial {Patterns} in the 2011 {London} {Riots}},
	volume = {7},
	issn = {1752-4512},
	url = {https://academic.oup.com/policing/article/7/1/21/1445612},
	doi = {10.1093/police/pas049},
	language = {en},
	number = {1},
	urldate = {2020-12-01},
	journal = {Policing: A Journal of Policy and Practice},
	author = {Baudains, Peter and Braithwaite, Alex and Johnson, Shane D.},
	month = mar,
	year = {2013},
	note = {Publisher: Oxford Academic},
	pages = {21--31}
}

@book{CoxStatisticalAnalysisSeries1966,
	series = {Methuen's {Monographs} on {Applied} {Probability} and {Statistics}},
	title = {The {Statistical} {Analysis} of {Series} of {Events}},
	isbn = {978-94-011-7803-7},
	url = {https://www.springer.com/gp/book/9789401178037},
	language = {en},
	urldate = {2020-12-01},
	publisher = {Springer Netherlands},
	author = {Cox, David R. and Lewis, P.A.W.},
	year = {1966}
}

@article{BartlettSpectralAnalysisPoint1963,
	title = {The {Spectral} {Analysis} of {Point} {Processes}},
	volume = {25},
	copyright = {© 1963 The Authors},
	issn = {2517-6161},
	url = {https://rss.onlinelibrary.wiley.com/doi/abs/10.1111/j.2517-6161.1963.tb00508.x},
	doi = {https://doi.org/10.1111/j.2517-6161.1963.tb00508.x},
	language = {en},
	number = {2},
	urldate = {2020-12-01},
	journal = {Journal of the Royal Statistical Society: Series B (Methodological)},
	author = {Bartlett, M. S.},
	year = {1963},
	pages = {264--281}
}

@article{BhardwajIndiasetsglobal2020,
	title = {India sets global record with single-day rise in coronavirus cases},
	url = {https://www.reuters.com/article/us-health-coronavirus-india-cases-idUSKBN25Q06A},
	language = {en},
	urldate = {2020-12-01},
	journal = {Reuters},
	author = {Bhardwaj, Mayank},
	month = aug,
	year = {2020}
}

@article{Donginteractivewebbaseddashboard2020,
	title = {An interactive web-based dashboard to track {COVID}-19 in real time},
	volume = {20},
	issn = {1473-3099, 1474-4457},
	url = {https://www.thelancet.com/journals/laninf/article/PIIS1473-3099(20)30120-1/abstract},
	doi = {10.1016/S1473-3099(20)30120-1},
	language = {English},
	number = {5},
	urldate = {2020-12-01},
	journal = {The Lancet Infectious Diseases},
	author = {Dong, Ensheng and Du, Hongru and Gardner, Lauren},
	month = may,
	year = {2020},
	pmid = {32087114},
	note = {Publisher: Elsevier},
	pages = {533--534}
}

@article{BrantinghamGangViolentCrime2020,
	title = {Is {Gang} {Violent} {Crime} {More} {Contagious} than {Non}-{Gang} {Violent} {Crime}?},
	issn = {0748-4518, 1573-7799},
	url = {http://link.springer.com/10.1007/s10940-020-09479-1},
	doi = {10.1007/s10940-020-09479-1},
	language = {en},
	urldate = {2021-02-24},
	journal = {Journal of Quantitative Criminology},
	author = {Brantingham, P. Jeffrey and Yuan, Baichuan and Herz, Denise},
	month = sep,
	year = {2020}
}

@article{LoefflerGunViolenceContagious2018,
	title = {Is {Gun} {Violence} {Contagious}? {A} {Spatiotemporal} {Test}},
	volume = {34},
	issn = {0748-4518, 1573-7799},
	shorttitle = {Is {Gun} {Violence} {Contagious}?},
	url = {http://link.springer.com/10.1007/s10940-017-9363-8},
	doi = {10.1007/s10940-017-9363-8},
	language = {en},
	number = {4},
	urldate = {2021-02-24},
	journal = {Journal of Quantitative Criminology},
	author = {Loeffler, Charles and Flaxman, Seth},
	month = dec,
	year = {2018},
	pages = {999--1017}
}

@article{PokharelProtestsIndiaNew2019,
	title = {Protests {Over} {India}’s {New} {Citizenship} {Law} {Widen}},
	url = {https://www.wsj.com/articles/protests-over-indias-new-citizenship-law-widen-11576501527},
	journal = {The Wall Street Journal},
	author = {Pokharel, Krishna and Purnell, Newley},
	year = {2019}
}

@misc{ACLEDMethodologyBriefCoronavirusRelated2020,
	title = {Methodology {Brief}: {Coronavirus}-{Related} {Events} in the {ACLED} {Dataset}},
	url = {https://acleddata.com/analysis/covid-19-disorder-tracker/},
	language = {en-US},
	urldate = {2021-02-20},
	author = {ACLED},
	month = mar,
	year = {2020}
}
\newpage

\appendix

\setcounter{equation}{0}
\setcounter{section}{0}
\setcounter{figure}{0}
\setcounter{table}{0}
\makeatletter
\renewcommand{\theequation}{S\arabic{equation}}
\renewcommand{\thefigure}{S\arabic{figure}}

\section{Appendix}

\subsection{ACLED event types}
\label{AppendixOne}
In this Appendix we illustrate the various categories the ACLED codebook uses to classify disorder events
\cite{ACLEDArmedConflictLocation2019}:

\vspace{0.2cm}
\noindent
{\bf{Violence against civilians}} involve one organized armed group deliberately inflicting violence against unarmed non-combatants. Perpetrators of violent acts can include state forces and affiliates, rebels, militias or other marginal subjects. Attempts to inflicting harm are also included, such as
attempted kidnappings.

\vspace{0.2cm}
\noindent
{\bf {Riots}} are characterized by demonstrators or mobs engaging in violent, disruptive actions such as property destruction. Riots can emerge from peaceful protests and are generally characterized by the use of unsophisticated weapons.

\vspace{0.2cm}
\noindent
{\bf{Protests}} refer to public demonstrations involving participants that do not engage in violent activity, although violence may be used against them. Symbolic acts 
such as publicly displaying flags are not coded as protests if they are not accompanied by a demonstration. Parliamentary walkouts and/or individual acts such as
self-harming are not included.

\vspace{0.2cm}
\noindent
{\bf {Battles}} involve violent interactions between politically organized armed groups at a particular time and location.  
At least two armed actors must be present; these may be armed and may include state, non-state and external entities.
There is no minimum threshold for the number of fatalities. 

\subsection{$k$-means clustering}
\label{AppendixTwo}

The purpose of $k$-means clustering is to partition a set of $n$ points
$\{x_{1},\cdots, x_{n}\}$ into $k$ clusters $C_{1}, \cdots , C_{k}$ 
\cite{HastieElementsStatisticalLearning2013a}. This iterative algorithm seeks to identify clusters
$C_i$ by considering their centroids
$\nu_i$ and by minimizing the average distance of the data points within it 
to the centroid. Therefore, the $k$-means algorithm tries to find
${\bf C} = \{C_{1}, \cdots , C_{k} \}$ and $\nu_i$ defined as

\begin{equation}
 \mathrm{\:\underset{\mathbf{C}}{arg \, min}}\sum_{i=1}^{k}\sum_{x\in C_i}\left \| x-\nu_i \right \|^{2}\
\end{equation}
Here, $\left \| x-\nu_i \right \|^{2}$ is the square of the Euclidean distance between the points in a given
cluster and its centroid $\nu_i$. Procedurally, $k$
centroids $\nu_i$ are initialized and each data point is assigned to
its closest centroid. The mean of the positions of all points within a cluster define the new centroid.
An iterative process ensues until discrepancies between iterations falls below a given threshold.
\subsubsection{Finding the optimal number of $k$}

To identify the optimal number of clusters $k^*$ we utilized the heuristic elbow method.
Here, $k$-means clustering is applied for several increasing values of $k$. Once clusters are identified, 
the sums of the square of the distance of each point within a cluster to its centroid is calculated. This $k$-dependent
quantity is termed WSS$(k)$, within-cluster sum. As $k$ increases,
more clusters are possible, hence, one may expect the WSS(k) to decrease as a function of $k$ as there may be a centroid closer
to them. However, beyond a critical value $k^*$ the decrease may be marginal, indicating that allowing for extra clusters
does not improve on the compactness of the clustering process. The value of $k^*$ beyond which
decreases in WSS asymptote yields the elbow, optimal value of $k^*$. 
In our work we use $1< k < 10$; as can be seen from 
for all three countries of interest, India, Israel and Mexico, the optimal $k^*$ value
is $k^* = 4$.

\begin{figure}[!t]
\centering
\includegraphics[scale=0.65]{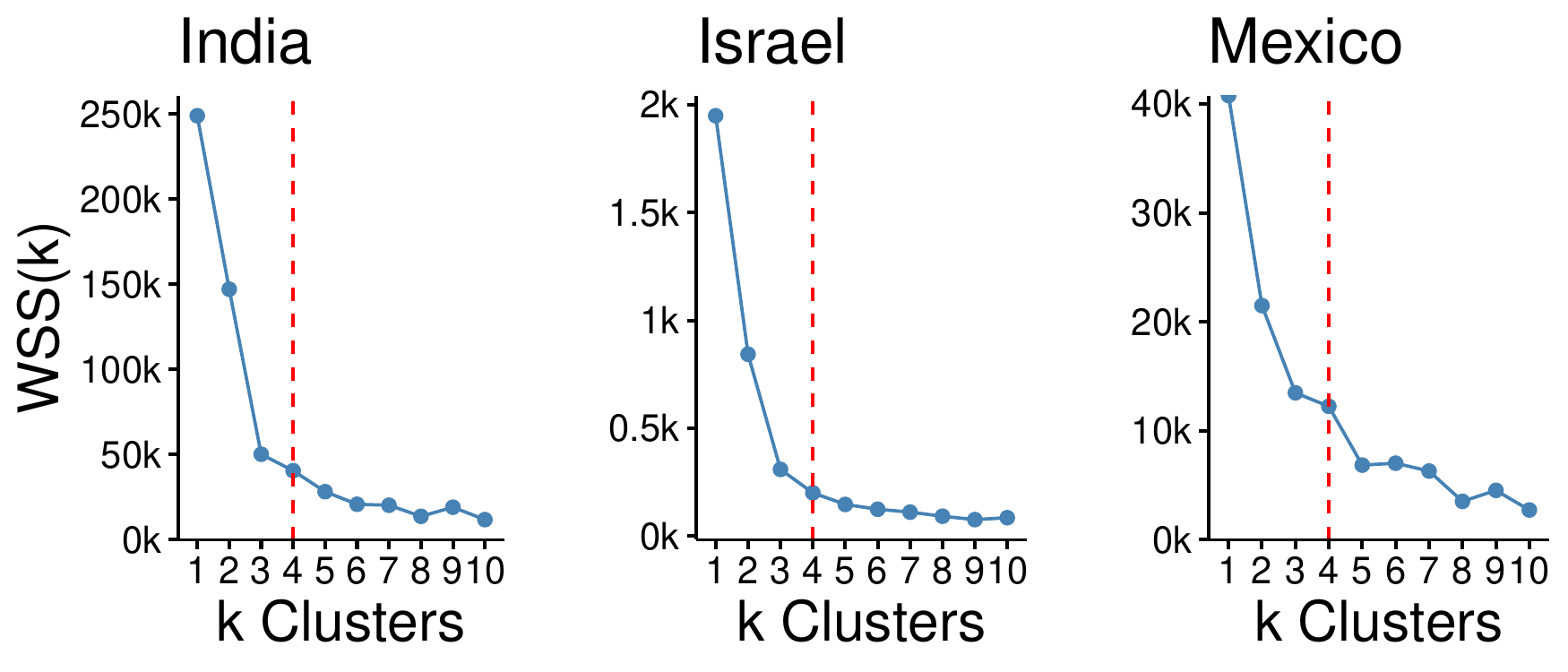}
\vspace{0.2cm}
\caption{From left to right: $k$-means clustering applied to Israel, India, Mexico. 
On the vertical axis is the WSS$(k)$. Note that the scales reflect the spatial extent
of the countries. India being the largest by territorial extent is associated to the largest WSS$(k)$ range,
India being the smallest is associated to the smallest WSS$(k)$ range.
The vertical
line denotes our elbow method best estimate for the optimal $k^*$ value which we identify
as $k^* =4$ in all countries.}
\label{fig:test}
\end{figure}

\subsection{Hawkes Process parameter estimation}
\label{AppendixThree}

We use MLE to derive the Hawkes process parameters
$\mu, \alpha, \beta$. These emerge as the ones that 
maximize the loglikelihood function defined as 

\begin{eqnarray}
    \mathrm{log}L(\mu, \alpha, \beta|t_1, ..., t_n) &=& \sum_{i=1}^n \log (\lambda(t_i)) - \int_{0}^{t_n }\lambda(t) dt \\
    \nonumber
    &=& \sum_{i=1}^{n}\mathrm{log}\left [\mu+\alpha\sum_{j=1}^{i-1}\mathrm{e}^{-\beta(t_i-t_j)} \right ]-\mu t_n+\frac{\alpha}{\beta}
    \sum_{i=1}^{n}\left [ \mathrm{e}^{-\beta(t_k-t_i)} -1\right ],
    \label{mle}
\end{eqnarray}
where $\{ t_1,...,t_n \}$ is the set of the times of occurrence of given events. 
The loglikelihood function compares the value of the intensity function of the Hawkes process
$\lambda(t)$ at event times $\{ t_1,...,t_n \}$ to the cumulative value of the function within the continuous 
interval $0 \leq t \leq t_n$. Maximizing the loglikelihood function yields parameters which best 
represent the actual event data. In this work we maximize  $\mathrm{log}L$  through the 
Nelder-Mead approach as available in the \texttt{ptproc} package in R  \cite{PengMultidimensionalPointProcess2002}.

\subsection{Event Distribution - Cluster wise}

\subsubsection{India}
\begin{figure}[!hbt] 
  \label{fig:sub-first}
  \centering
  \includegraphics[width=.5\linewidth]{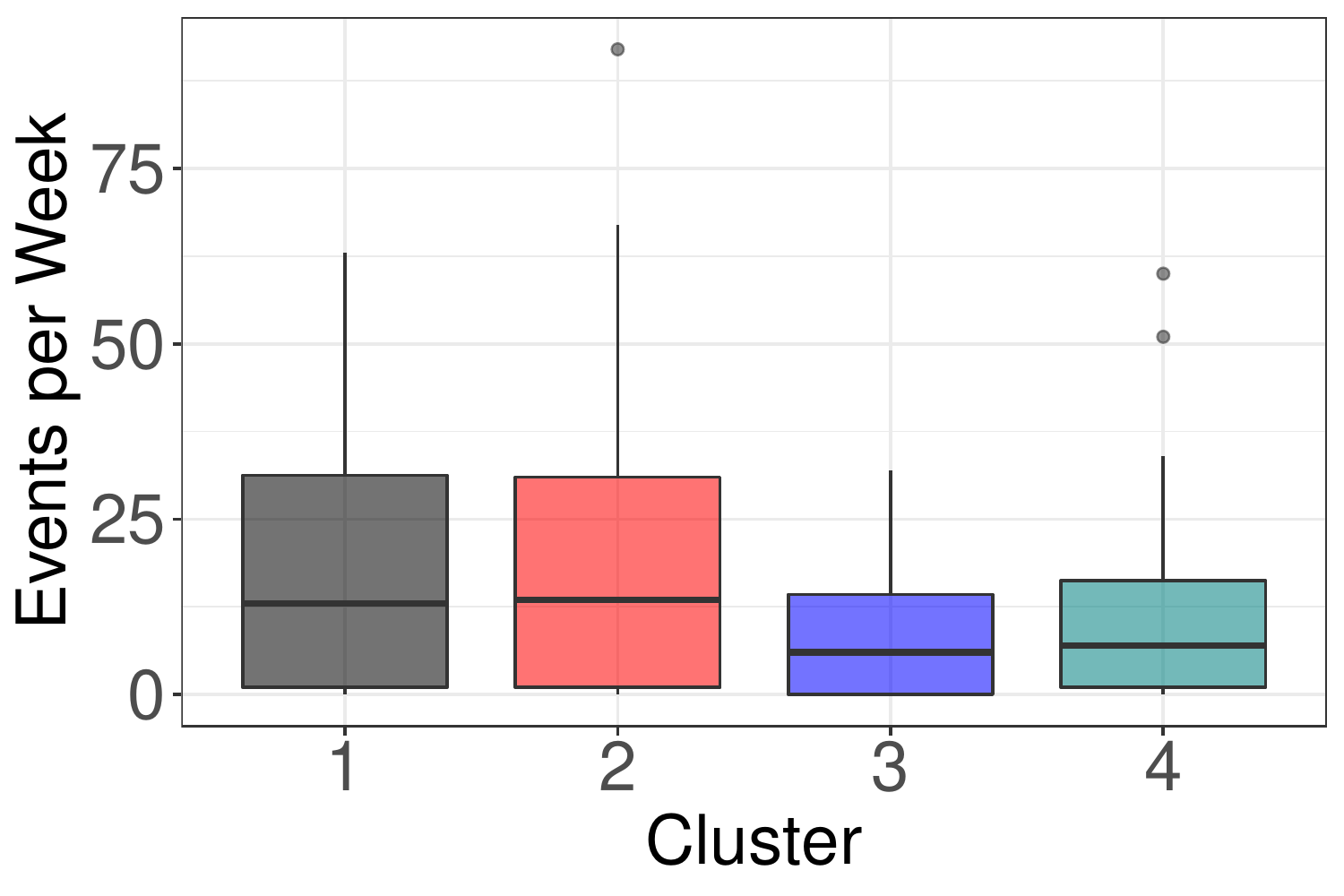}  
  \caption{Cluster-wise boxplot of disorder events in India. The most occurrences arise in clusters C1, C2, where the
  most densely populated states are located. C4 displays several outliers.}
\label{fig:indiats2}
  \end{figure}

Distribution summaries are shown in 
Fig.\,\ref{fig:indiats2}: C2 has the highest average number of disorders per week 
and the highest variability, followed by C1. Interestingly, while C4 has the second-lowest average number of disorders, 
it exhibits outliers, coinciding with week $j=19$ (51 events) and week $j=24$ (60 events).

\subsubsection{Israel}

\begin{figure}[!h] 
  \label{fig:sub-first}
  \centering
  \includegraphics[width=.5\linewidth]{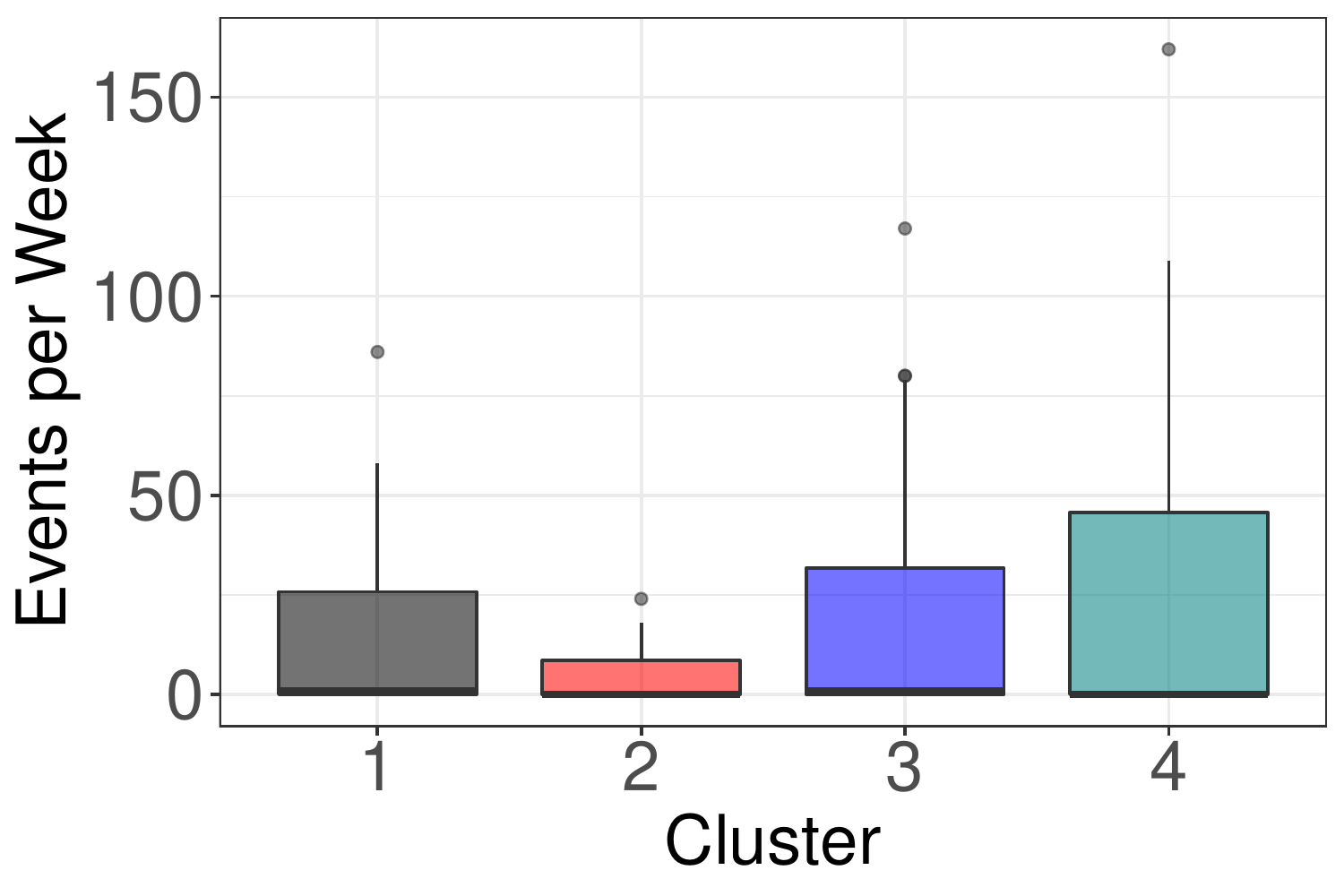}  
  \caption{Cluster-wise boxplot of disorder events in Israel. The most occurrences arise in clusters C1, C3 and C4 where the
  major cities of Haifa, Tel Aviv and Jerusalem are located.}
\label{fig:israts2}
  \end{figure}

Figure \ref{fig:israts2} reveals low values of averaged weekly disorders, 
however many outliers emerge corresponding to the interval between weeks $j=37$ and $j=50$ mentioned above. 
\vspace{0.5cm}  
  
\subsubsection{Mexico}

\begin{figure}[!hbt] 
  \label{fig:sub-first}
  \centering
  \includegraphics[width=.5\linewidth]{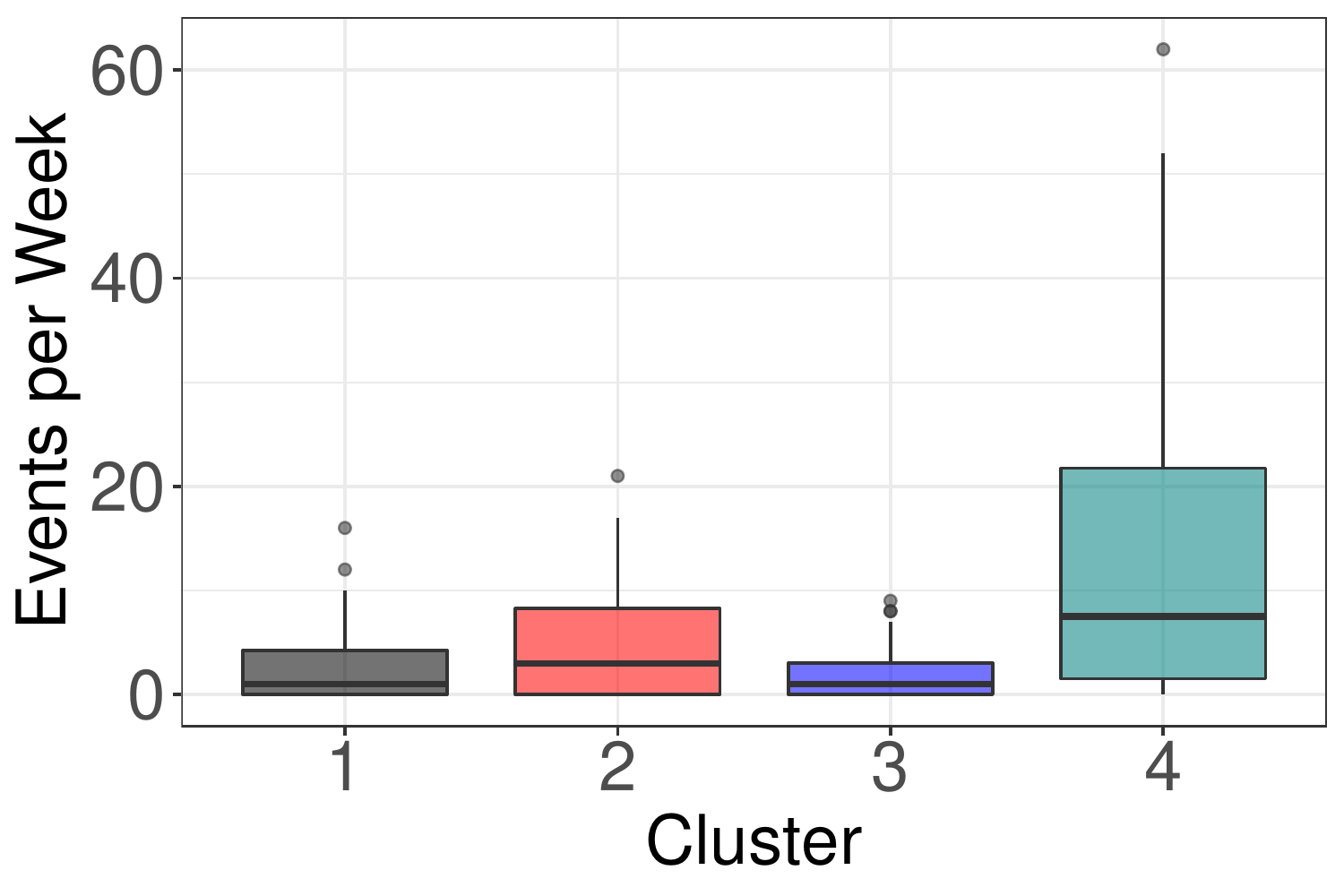}  
  \caption{Cluster-wise boxplot of disorder events in Mexico. The most occurrences arise in cluster C4, 
  where the most populous and dense areas of Mexico City and Mexico state are located.}
\label{fig:mexicots2}
  \end{figure}

Figure \ref{fig:mexicots2} summarizes the distribution of events in Mexico at the weekly level. As mentioned, C4 has the highest average and variability in event counts, 
followed by C2, whereas in C3 and C1 fewer events are recorded. Interestingly,
C1 is characterized by a very low variability. Thus, while spikes in activity and fluctuations emerge in other clusters, events in C1 are more uniformly distributed.
\newpage

\subsection{Cluster-based analysis: Pearson's correlation coefficients}
\label{AppendixFour}

In this section we list the numerical values of the Pearson coefficient $r$ correlating the 
number of weekly of events in pairs of clusters within a given country.
If we denote  two clusters within a country C$_X$ and C$_Y$  then $r$ is defined as

\begin{eqnarray}
r = \frac {\mathrm E(X - \mu_x)E(Y - \mu_y)}{\sigma_X \sigma_Y}
\end{eqnarray}
where $X,Y$ are the sets of weekly data in clusters C$_X$ and C$_Y$,  respectively, 
$\mu_X, \mu_Y$ their averages, and $\sigma_X, \sigma_Y$ their standard deviations.
Pearson correlation coefficient ranges from $−1$ to $1$.  $r=1$ 
implies a perfect, positive, linear relationship between the two datasets whereas
$r=-1$ implies a perfect negative one. As $|r|$ decreases, 
correlations become weaker, so that $r = 0$ implies data points in the two sets $X,Y$ are not correlated.
In our work, $X,Y$ are the either the sets of weekly events $\{n^X_j\}, \{n^Y_j \}$ in each cluster 
or the sets of differentiated weekly events  $\{\Delta n^X_j\}, \{\Delta n^Y_j \}$ where  
$\Delta n^X_j = n^X_{j}-n^X_{j-1}$ and $\Delta n^Y_j = n^Y_{j}-n^Y_{j-1}$. Below we show how these quantities
manifest in each of the three countries under investigation.
\newpage
\subsubsection{India}
\begin{table}[!hbt]
\footnotesize
\centering
\begin{tabular}{l|llll}
\hline
& C1 & C2 & C3 & C4 \\ \hline
C1 & 1.000 &  &  &  \\
C2 & 0.678 & 1.000 &  &  \\
C3 & 0.724 & 0.595 & 1.000 &  \\
C4 & 0.598 & 0.644 & 0.658 & 1.000 \\ \hline
\end{tabular}
\end{table}
\begin{table}[!hbt]
\footnotesize
\centering
\begin{tabular}{l|llll}
\hline
 & C1 & C2 & C3 & C4 \\ \hline
C1 & 1.000 &  &  &  \\
C2 & 0.124 & 1.000 &  &  \\
C3 & 0.219 & 0.301 & 1.000 &  \\
C4 & 0.450 & 0.339 & 0.512 & 1.000 \\ \hline
\end{tabular}
\vspace{0.4cm}
\caption{Pearson's correlation matrices for India and shown in Fig. 6.
Top: Entries represent
correlation coefficients $r$ derived on weekly events $\{ n_j\}$ for the period January 3$^{\rm rd}$
to December 12$^{\rm th}$ 2020 and between the associated clusters.
Overall, correlation values are moderately large and uniform. The highest $r = 0.724$ is observed between clusters C1 and C3.
Bottom: Entries represent correlation coefficients $r$ derived on differentiated weekly events $\{ \Delta n_j\}$ and
show much weaker correlation, implying a reduced synchrony in the rate of change of the occurrence of events.}
\end{table}

\subsubsection{Israel}
\footnotesize
\begin{table}[!hbt]
\footnotesize
\centering
\begin{tabular}{l|llll}
\hline
 & C1 & C2 & C3 & C4 \\ \hline
C1 & 1.000 &  &  &  \\
C2 & 0.996 & 1.000 &  &  \\
C3 & 0.998 & 0.995 & 1.000 &  \\
C4 & 0.998 & 0.995 & 0.999 & 1.000 \\ \hline
\end{tabular}
\end{table}
\begin{table}[!hbt]
\footnotesize
\centering
\begin{tabular}{l|llll}
\hline
 & C1 & C2 & C3 & C4 \\ \hline
C1 & 1.000 &  &  &  \\
C2 & 0.972 & 1.000 &  &  \\
C3 & 0.986 & 0.954 & 1.000 &  \\
C4 & 0.986 & 0.958 & 0.995 & 1.000 \\ \hline
\end{tabular}
\vspace{0.4cm}
\caption{Pearson's correlation matrices for Israel and shown in Fig. 10. Top: Entries represents 
correlation coefficients $r$ derived on weekly events $\{ n_j\}$ for the period January 3$^{\rm rd}$
to December 12$^{\rm th}$ 2020 and between the associated clusters.
Correlation values approach unity, revealing large synchrony within the country.
Bottom: Entries represent correlation coefficients $r$ derived on differentiated weekly events $\{ \Delta n_j\}$. These remain
very large, confirming the large degree of synchrony in the rate of change of events in the country.}
\end{table}
\newpage
\subsubsection{Mexico}

\begin{table}[!hbt]
\footnotesize
\centering
\begin{tabular}{l|llll}
\hline
 & C1 & C2 & C3 & C4 \\ \hline
C1 & 1.000 &  &  &  \\
C2 & 0.675 & 1.000 &  &  \\
C3 & 0.632 & 0.571 & 1.000 &  \\
C4 & 0.753 & 0.826 & 0.772 & 1.000 \\ \hline
\end{tabular}
\end{table}
\begin{table}[!hbt]
\footnotesize
\centering
\begin{tabular}{l|llll}
\hline
 & C1 & C2 & C3 & C4 \\ \hline
C1 & 1.000 &  &  &  \\
C2 & -0.092 & 1.000 &  &  \\
C3 & 0.203 & -0.445 & 1.000 &  \\
C4 & 0.286 & 0.140 & 0.246 & 1.000 \\ \hline
\end{tabular}
\vspace{0.4cm}
\caption{Pearson's correlation matrices for Mexico and shown in Fig. 14. Top: Entries represents 
correlation coefficients $r$ derived on weekly events $\{ n_j\}$ for the period January 3$^{\rm rd}$
to December 12$^{\rm th}$ 2020 and between the associated clusters.
Overall, correlation values are moderately large. The highest $r = 0.826$ is observed between the geographically 
contiguous clusters C2 and C4. The lowest $r = 0.571$ is observed between clusters C2 and C4. 
Bottom: Entries represent correlation coefficients $r$ derived on differentiated weekly events $\{ \Delta n_j\}$
show vanishing or even negative correlation and implying lack of synchrony in the rate of change of the occurrence of events.}
\end{table}

\subsection{Hawkes process in a restricted time window}
\label{AppendixFive}

\normalsize
In this section we apply the Hawkes process to disorder events recorded from the CDT from January 3${\rm rd}$ to October 10$^{\rm th}$ 2020. 
Similarly to what observed for the entire data set, the Hawkes process outperforms the Poisson process in all three countries and in all clusters,  even in this limited time range. A noteworthy observation is that while the sequence of events in C4 in Israel is appropriately described by a Hawkes process until October 10$^{\rm th}$ 2020 as per Table\,\ref{israelappendix}, the sequence of events that extends to December 12$^{\rm th}$ is not as per Table 4, confirming that 
disorders in Israel in Fall 2020 are even extremely clustered than what predicted by Hawkes processes.  
\newpage
\subsubsection{India}
\begin{table}[!hbt]
\footnotesize
\centering
\begin{tabular}{c|l|ccccc}
\hline
\multicolumn{2}{c|}{\textbf{Cluster}} & \textbf{\begin{tabular}[c]{@{}c@{}}India\\ (all)\end{tabular}} & \textbf{\begin{tabular}[c]{@{}c@{}}India\\ (C1)\end{tabular}} & \textbf{\begin{tabular}[c]{@{}c@{}}India\\ (C2)\end{tabular}} & \textbf{\begin{tabular}[c]{@{}c@{}}India\\ (C3)\end{tabular}} & \textbf{\begin{tabular}[c]{@{}c@{}}India\\ (C4)\end{tabular}} \\ \hline
 & Number of events & 2,744 & 852 & 946 & 408 & 538 \\ \cline{2-7} 
 & $\mu$ & 0.291 & 0.537 & 0.332 & 0.120 & 0.662 \\
 & $\alpha$ & 2.075 & 1.447 & 1.538 & 0.495 & 1.518 \\
 & $\beta$ & 2.020 & 1.223 & 1.400 & 0.462 & 1.078 \\ \cline{2-7} 
 & $\gamma$ & 0.973 & 0.845 & 0.910 & 0.933 & 0.710 \\
 & $\mu/(1-\gamma)$ & 10.777 & 3.464 & 3.666 & 1.791 & 2.282 \\
 & Hawkes AIC & -9230 & -825 & -1274 & 204 & -89 \\
 & Poisson AIC & -7360 & -371 & -526 & 428 & 195 \\ \cline{2-7} 
 & KS Stat, $D$ & 0.147 & 0.097 & 0.103 & 0.154 & 0.063 \\
 & KS Crit 95\%, $D^{95}_{\rm c}$ & 0.161 & 0.118 & 0.145 & 0.246 & 0.113 \\
 & KS Crit 99\%, $D^{99}_{\rm c}$ & 0.193 & 0.141 & 0.174 & 0.295 & 0.135 \\ \hline
\end{tabular}
\vspace{0.4cm}
\caption{Statistical outcomes of the Hawkes process applied to data from India up to October 10$^{\rm th}$ 2020.
The Hawkes process outperforms the baseline Poisson process both nationwide and in each
cluster, since the Hawkes AIC is always less than the Poisson AIC. The Hawkes process 
passes the KS test at the 95\% significance level in all cases.}
\end{table}

\subsubsection{Israel}

\begin{table}[!hbt]
\footnotesize
\centering
\begin{tabular}{c|l|ccccc}
\hline
\multicolumn{2}{c|}{\textbf{Cluster}} & \textbf{\begin{tabular}[c]{@{}c@{}}Israel\\ (all)\end{tabular}} & \textbf{\begin{tabular}[c]{@{}c@{}}Israel\\ (C1)\end{tabular}} & \textbf{\begin{tabular}[c]{@{}c@{}}Israel\\ (C2)\end{tabular}} & \textbf{\begin{tabular}[c]{@{}c@{}}Israel\\ (C3)\end{tabular}} & \textbf{\begin{tabular}[c]{@{}c@{}}Israel\\ (C4)\end{tabular}} \\ \hline
 & Number of events & 1,197 & 285 & 76 & 373 & 463 \\ \cline{2-7} 
 & $\mu$ & 0.341 & 0.207 & 0.640 & 0.184 & 0.081 \\
 & $\alpha$ & 20.927 & 10.383 & 6.865 & 11.159 & 13.901 \\
 & $\beta$ & 19.749 & 8.987 & 5.368 & 10.107 & 13.626 \\ \cline{2-7} 
 & $\gamma$ & 0.944 & 0.866 & 0.782 & 0.906 & 0.980 \\
 & $\mu/(1-\gamma)$ & 6.089 & 1.544 & 2.935 & 1.957 & 4.050 \\
 & Hawkes AIC & -7871 & -742 & -66 & -1366 & -2290 \\
 & Poisson AIC & -1759 & 375 & 6 & 312 & 200 \\ \cline{2-7} 
 & KS Stat, $D$ & 0.104 & 0.164 & 0.122 & 0.131 & 0.257 \\
 & KS Crit 95\%, $D^{95}_{\rm c}$ & 0.164 & 0.207 & 0.375 & 0.224 & 0.355 \\
 & KS Crit 99\%, $D^{99}_{\rm c}$ & 0.196 & 0.249 & 0.449 & 0.268 & 0.381 \\ \hline
\end{tabular}
\vspace{0.4cm}
\caption{Statistical outcomes of the Hawkes process applied to data from Israel up to October 10$^{\rm th}$ 2020.
The Hawkes process outperforms the baseline Poisson process both nationwide and in each cluster, since the Hawkes AIC is always less than the Poisson AIC. The Hawkes process 
passes the KS test at the 95\% significance level in all cases.}
\label{israelappendix}
\end{table}

\newpage
\subsubsection{Mexico}

\begin{table}[!hbt]
\footnotesize
\centering
\begin{tabular}{c|l|ccccc}
\hline
\multicolumn{2}{c|}{\textbf{Cluster}} & \textbf{\begin{tabular}[c]{@{}c@{}}Mexico\\ (all)\end{tabular}} & \textbf{\begin{tabular}[c]{@{}c@{}}Mexico\\ (C1)\end{tabular}} & \textbf{\begin{tabular}[c]{@{}c@{}}Mexico\\ (C2)\end{tabular}} & \textbf{\begin{tabular}[c]{@{}c@{}}Mexico\\ (C3)\end{tabular}} & \textbf{\begin{tabular}[c]{@{}c@{}}Mexico\\ (C4)\end{tabular}} \\ \hline
 & Number of events & 1,193 & 135 & 254 & 91 & 703 \\ \cline{2-7} 
 & $\mu$ & 1.330 & 0.460 & 0.651 & 0.143 & 0.985 \\
 & $\alpha$ & 2.968 & 2.911 & 1.845 & 0.159 & 2.496 \\
 & $\beta$ & 2.287 & 1.085 & 0.906 & 0.110 & 1.782 \\ \cline{2-7} 
 & $\gamma$ & 0.771 & 0.373 & 0.491 & 0.695 & 0.714 \\
 & $\mu/(1-\gamma)$ & 5.807 & 0.733 & 1.278 & 0.468 & 3.444 \\
 & Hawkes AIC & -2337 & 318 & 325 & 315 & -659 \\
 & Poisson AIC & -1781 & 356 & 386 & 330 & -324 \\ \cline{2-7} 
 & KS Stat, $D$ & 0.036 & 0.065 & 0.071 & 0.134 & 0.029 \\
 & KS Crit 95\%, $D^{95}_{\rm c}$ & 0.081 & 0.147 & 0.116 & 0.275 & 0.096 \\
 & KS Crit 99\%, $D^{99}_{\rm c}$ & 0.097 & 0.176 & 0.139 & 0.330 & 0.115 \\ \hline
\end{tabular}
\vspace{0.4cm}
\caption{Statistical outcomes of the Hawkes process applied to data from Mexico up to October 10$^{\rm th}$ 2020.
The Hawkes process outperforms the baseline Poisson process both nationwide and in each 
cluster, since the Hawkes AIC is always less than the Poisson AIC. The Hawkes process 
passes the KS test at the 95\% significance level in all cases.}
\end{table}

\end{document}